\documentclass{aa}  

\usepackage{graphicx, caption, subcaption}
\usepackage{txfonts}
\usepackage{siunitx}
\begin{document}

   \title{Rotational spectroscopy of \textit{n}-propanol: \textit{Aa} and \textit{Ag} conformers\thanks{The supplementary material, our predictions of the rotational spectra (\textit{*.cat} files), are only available at CDS.}}


   \author{O. Zingsheim\inst{1}
          \and
          J. Maßen\inst{1}
          \and
          H.~S.~P. M\"uller\inst{1}
          \and
          B. Heyne\inst{1}
          \and
          M. Fatima\inst{1}
          \and
          L. Bonah\inst{1}
          \and
          A. Belloche\inst{2}
          \and
          F. Lewen\inst{1}
          \and
          S. Schlemmer\inst{1}\fnmsep
          }

   \institute{I. Physikalisches Institut, Universit\"at zu K\"oln, Z\"ulpicher Stra{\ss}e 77, 50937 K\"oln, Germany\\
              \email{zingsheim@ph1.uni-koeln.de;~hspm@ph1.uni-koeln.de}
         \and
             Max-Planck-Institut für Radioastronomie, Auf dem Hügel 69, 53121 Bonn, Germany\\
             }
\date{Received 2022, A\&A accepted \href{https://doi.org/10.1051/0004-6361/202243571}{doi:10.1051/0004-6361/202243571}}



  \abstract
   {The primary alcohol \textit{n}-propanol (i.e., normal-propanol or propan-1-ol; C$_3$H$_7$OH) occurs in five different conformers: \textit{Ga}, \textit{Gg}, \textit{Gg'}, \textit{Aa}, and \textit{Ag}.
All rotational spectra of the three conformers of the \textit{G} family are well described, making astronomical search of their spectroscopic signatures possible, as opposed to those of the \textit{Aa} and \textit{Ag} conformers.}
   {Our goal is to facilitate the astronomical detection of \textit{Aa} and \textit{Ag} conformers of \textit{n}-propanol by characterizing their rotational spectra.}
   {We recorded the rotational spectra of \textit{n}-propanol in the frequency domain of 18$-$505\,GHz.
   Additional double-modulation double-resonance (DM-DR) measurements were performed, more specifically with the goal to unambiguously assign weak transitions of the \textit{Aa} conformer and to verify assignments of the \textit{Ag} conformer.}
   {We derived a spectroscopic quantum mechanical model with experimental accuracy (with $J_\textrm{max}=70$ and $K_{a,\textrm{max}}=6$) for \textit{Aa n}-propanol. Furthermore, we unambiguously assigned transitions (with $J_\textrm{max}=69$ and $K_{a,\textrm{max}}=9$) of \textit{Ag n}-propanol; in doing so, we prove the existence of two tunneling states, \textit{Ag}$^+$ and \textit{Ag}$^-$.}
   {The astronomical search of all five conformers of \textit{n}-propanol is now possible via their rotational signatures. These are applied in a companion article on the detection of \textit{n}-propanol toward the hot molecular core Sgr B2(N2).}
   
   \keywords{ molecular data -- methods: laboratory: molecular -- techniques: spectroscopic -- ISM: molecules  -- radio lines: ISM -- submillimeter: ISM 
               }
   \maketitle
%

\section{Introduction}

To date, more than 260 molecules have been detected in space.\footnote{\url{https://cdms.astro.uni-koeln.de/classic/molecules}}
Many rotational molecular spectra are collected in databases such as the Cologne Database for Molecular Spectroscopy (CDMS) \citep{CDMS_2001,CDMS_2005,CDMS_2016},  allowing astronomers to search for their spectroscopic signatures in space.
Several complex organic molecules (COMs) have been found thus far, for instance, in the warm gas around protostars in star forming regions, such as formamide \citep{1971ApJ...169L..39R}, ethyl cyanide \citep{1977ApJ...218..370J}, and ethanediol \citep{Hollis_2002}.
One prominent example is the Protostellar Interferometric
Line Survey  \citep[PILS;][]{PILS_2016} with the Atacama Large Millimeter/submillimeter Array (ALMA). 
Among the COMs, alcohols have also been found in space.
So far, methanol \citep[CH$_3$OH;][]{1970ApJ...162L.203B} and ethanol \citep[C$_2$H$_5$OH;][]{ethanol_anti_1975,ethanol_gauche_1997} have been detected. Ethanol occurs as two conformers, so-called anti-ethanol \citep{ethanol_anti_1975} and gauche-ethanol \citep{ethanol_gauche_1997}, and both have been referenced based on their first detection.
The next heavier alcohol, propanol (C$_3$H$_7$OH), is thus another candidate for a detection in space, particularly in environments with large amounts of ethanol.
The possible detection of propanol may be further supported by the detection of its isomer, ethyl methyl ether (or methoxyethane, CH$_3$CH$_2$OCH$_3$) in space \citep{ethyl_methyl_ether_2015,ethyl_methyl_ether_2018,ethyl_methyl_ether_2020}.

Propanol itself occurs in two isomers, as a primary alcohol \textit{n}-propanol (normal-propanol, propan-1-ol, normal propyl alcohol; CH$_3$CH$_2$CH$_2$OH), where the hydroxyl group \mbox{($-$OH)} is attached to the outer carbon atom, and as a secondary alcohol \textit{i}-propanol (iso-propanol, propan-2-ol, or iso-propyl alcohol; CH$_3$CH(OH)CH$_3$), where the hydroxyl group is attached to the central carbon atom.
\textit{n}-Propanol, the subject of this study, occurs in five different conformers: \textit{Ga}, \textit{Gg}, \textit{Gg'}, \textit{Aa}, and \textit{Ag}, see  \citet{Propanol_U2_1968,Propanol_U5_1970_Abdurakhmanov_123,Viniti_Ag_1976,n-propanol_DreizlerScappini_1981,LOTTA_n-propanol,n-propanol_Maeda_2006,n-propanol_Kisiel_2010} and references therein.
The capital letters \textit{G} and \textit{A} refer to the rotation of the heavy nuclei plane \mbox{C-C-C} compared to \mbox{C-C-O}. An  "anti" (\textit{A}) configuration\footnote{The term "trans"\ (\textit{T}) was often used instead of "anti"\ historically, although it should only be used to refer to planar molecules.}
describes a rotation by 180\,$^\circ$ and "gauche"\ (\textit{G}) by about 60\,$^\circ$.
The small letters refer to the rotation of the dihedral angle of the hydroxy group (-OH).

The \textit{G} family of conformers (\textit{Ga}, \textit{Gg}, and \textit{Gg'}) of \textit{n}-propanol has been rather extensively studied in rotational spectroscopy, as in, for instance,~\citet{Propanol_U5_1970_Abdurakhmanov_123,n-propanol_Maeda_2006,n-propanol_Kisiel_2010} and references therein.
A single state analysis of \textit{Ga} was able to fit 2865 lines up to 375\,GHz, but systematic deviations occurred for some transitions with higher $J$ quantum numbers \citep{n-propanol_Maeda_2006}. 
Based on this observation, a combined analysis of the three \textit{G} conformers was carried out by taking into account Coriolis interaction and Fermi resonances \citep{n-propanol_Kisiel_2010}. Thanks to the quantum mechanical models of rotational spectra with experimental accuracy \citep{n-propanol_Kisiel_2010}, conformers of the \textit{G} family can be found in space.
Furthermore, this distinctive analysis gives confidence that the \textit{G} and \textit{A} conformers can be treated separately as within the \textit{G} family, the conformers seem to be unaffected by conformers of the \textit{A} family (\textit{Aa} and \textit{Ag}).

The ab initio calculations have shown some scatter for the relative energies of \textit{n}-propanol conformers, as, for instance, in ~\citet{LOTTA_n-propanol,Propanol_rel-energies_2005} and references therein (a detailed overview is given in \citet{n-propanol_Maeda_2006}, but they agree in the fact that all conformers, in particular from both \textit{G} and \textit{A} families, are rather close in energy).\footnote{For instance, the \textit{Ga} conformer was predicted
to be the energetically lowest energy conformer with the others higher in energy by $E(Aa)=24$\,cm$^{-1}$ (35\,K),
$E(Gg) =35$\,cm$^{-1}$ (50\,K), $E(Ag) =42$\,cm$^{-1}$ (60\,K), and $E(Gg')=49$\,cm$^{-1}$ (70\,K) \citep{Propanol_rel-energies_2005}. Experimentally derived values for the \textit{G} family are $E(Ga)=0\,$cm$^{-1}$, $E(Gg)=47.8$\,cm$^{-1}$ (69\,K), and $E(Gg')=50.8$\,cm$^{-1}$ (73\,K) \citep{n-propanol_Kisiel_2010}.}
Considering only Boltzmann statistics, the \textit{Aa} and \textit{Ag} conformers are expected to have about the same chance of being detected in the warm interstellar medium as the \textit{G} family conformers.

\begin{figure}[t]
\centering
\includegraphics[width=1.0\linewidth]{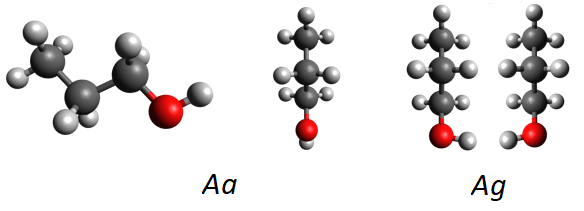} 
\caption{\textit{Aa} and \textit{Ag} conformers of \textit{n}-propanol. Hydrogen, carbon, and oxygen atoms are depicted with white, black, and red spheres, respectively.
The atoms C-C-C-O-H span a (symmetry) plane for the \textit{Aa} conformer ($C_s$ symmetry) and it is shown as face-on and side view for clarification. Then, 
\textit{Ag n}-propanol occurs in two mirror-image forms leading to the \textit{Ag}$^+$ and \textit{Ag}$^-$ tunneling states.
The \textit{G} conformers are not depicted here, but the oxygen is out of the heavy nuclei plane for all of them.}
\label{Fig1:T_conformers}
\end{figure}

The most advanced rotational analysis of the \textit{Aa n}-propanol was derived by using a Fourier transform microwave (FTMW) spectrometer (8$-$18\,GHz) and a Stark modulated microwave spectrometer (up to 40\,GHz) \citep{n-propanol_DreizlerScappini_1981}.
Four conformers of \textit{n}-propanol occur as mirror configurations (two-fold degeneracy due to "left-" and "right-handed" versions, $C_1$ symmetry), except for the \textit{Aa} conformer, which has a plane of symmetry and one-fold degeneracy ($C_s$ symmetry), as shown in Fig.~\ref{Fig1:T_conformers}.
The \textit{Ag} conformer occurs in mirror images (see Fig.~\ref{Fig1:T_conformers}), however, the barrier for hydrogen tunneling between the two equivalent \textit{Ag} forms is low enough to distinguish symmetric
\textit{Ag}$^+$ and antisymmetric \textit{Ag}$^-$ tunneling states \citep{Viniti_Ag_1976,n-propanol_Kisiel_2010}.
The tunneling transition state has $C_s$ symmetry and tunneling-rotation interactions need to be taken into account, as has been done, for instance, for gauche-ethanol \citep{PEARSON2008394} and gauche-propanal more recently \citep{Propanal_2_Pickett_Scroggin,ZINGSHEIM2017_Propanal,ZINGSHEIM2021_Propanal}.
So far, only eleven doublet features up to 30\,GHz were tentatively assigned to \textit{Ag$^+$} and \textit{Ag$^-$} tunneling states with $J_{\textrm{max}}=4$ \citep{Viniti_Ag_1976}.

The main aim of the present study is to facilitate the astronomical detection of \textit{Aa n}-propanol and \textit{Ag n}-propanol.
For this purpose, the rotational spectra are measured in selected frequency ranges from 18 to 505\,GHz. The employed experimental setups are presented in Sect.~\ref{Sec:Experiment}.
The experimental data are analyzed in Sect.~\ref{Sec:ExperimentalResults} with a focus on secure assignments of the perturbed systems.
Next, the spectroscopic results and the astronomical search of \textit{n}-propanol are discussed in Sect.~\ref{Sec:Discussion}.
Finally, our main conclusions are given in Sect.~\ref{Sec:Conclusion}.

\section{Experimental techniques}     
\label{Sec:Experiment}                

We recorded rotational spectra of \textit{n}-propanol (i) in the 17.5$-$26.5\,GHz region using chirped-pulse Fourier transform microwave (CP-FTMW) jet spectroscopy and (ii) in the regions of 33$-$67\,GHz, 70$-$129\,GHz, and 160$-$505\,GHz using conventional $2f$ demodulation absorption techniques.
Additional (iii) double-modulation double-resonance (DM-DR) measurements support assignments of weaker transitions to the \textit{Aa} conformer and, in particular, secure the identification of various $^qR$ series\footnote{Transitions with identical selection rules are called a series, here abbreviated as $^{\Delta K_a} \Delta J$, with $\Delta J$ represented by capital letters $P$, $Q$, and $R$ for $\Delta J=-1,\pm0$ or $+1$, respectively, and $\Delta K_a=-1,\pm0$, or $+1$ by superscripts $p$, $q$, or $r$, respectively. For instance, $a$-type transitions ($\Delta K_a=0$) with $\Delta J=+1$ are called $^qR$ series transitions. A subscript may be used additionally to highlight the respective $K_a$ quantum number ($^qR_{K_a}$ series).} of the \textit{Ag} conformer.
The (i) CP-FTMW, (ii) conventional \mbox{(sub-)}millimeter wave (MMW) absorption spectrometers, and (iii) the DM-DR setup are introduced in detail in (i) \citet{cp-FTMW-cologne},~(ii) \citet{Ordu2019_Acetone,Drumel2015_OSSO}, and (iii) \citet{Zingsheim2021_DMDR}, respectively.
Thus, their main characteristics are only briefly summarized in the following.

In terms of (i) 
the 17.5$-$26.5\,GHz range was covered in steps of 1\,GHz bandwidth using a CP-FTMW spectrometer.
In contrast to the above-described setup \citep{cp-FTMW-cologne}, mixing up the CP and mixing down the molecular signal is no longer required as the CP is generated directly in the target frequency region using an arbitrary waveform generator (Keysight M8195A 65 GSa/s).
In the present setup, each 1\,GHz broad CP spanned 2\,\textmu s and was amplified with a 4\,W amplifier.
These amplified pulses were then broadcast via a horn antenna to excite the molecular ensemble.
After the pulse was stopped, the resulting free induction decay (FID) of the molecules was collected with another horn antenna, amplified using a low noise amplifier, and recorded for 10\,\textmu s on a 100\,GS/s oscilloscope.
Each measurement consists of 65\,000-100\,000 acquisitions, which were performed with a 10\,Hz repetition rate.
The averaged FIDs were then Fourier transformed using a Hanning window, which resulted in a frequency resolution of about 290\,kHz, considering the full width at half maximum (FWHM).
The sample was prepared by mixing 1\,mL of liquid \textit{n}-propanol (Sigma-Aldrich, 97.7\,\% purity) with 7\,bar neon in a 10\,L container.
The mixture was supersonically expanded with a relative stagnation pressure of 1\,bar and an opening duration of 500\,\textmu s.

For (ii) either an Agilent E8257D or a Rohde \& Schwarz SMF100A synthesizer generated electromagnetic waves.
Electromagnetic waves were directly coupled to a horn antenna to record single spectra in the 33$-$67\,GHz region.
We used in-house developed electronics for operating an amplifier tripler chain in full saturation mode to fully cover the 70$-$129\,GHz region and used commercially available cascaded doublers and triplers from Virginia Diodes Inc. (VDI) to fully cover the 160$-$505\,GHz region.
Frequency modulation (FM) increases the signal-to-noise ratio (S/N)
and the 2$f$ demodulation creates absorption features that are close to the second derivative of a Voigt profile. We used a modulation frequency of $f\approx47$\,kHz and its amplitude is on the order of the FWHM of the absorption profiles.
The absorption path was 14\,m in a single pass configuration for measurements up to 130\,GHz \citep{Ordu2019_Acetone} and 10\,m in double pass configuration otherwise \citep{Drumel2015_OSSO}. Here,
\textit{n}-Propanol was vaporized into the cell to pressures in the range of 10$-$40\,\textmu bar.

For (iii) we used both synthesizers, namely, Agilent as the so-called probe and Rohde \& Schwarz as the pump, simultaneously for DM-DR measurements; a modulated pump radiation was applied in addition to the frequency modulated probe radiation.
The probe signal was usually frequency tripled, and in-house developed electronics used for operation in full saturation mode ensured a stable output power of about 1\,mW (0\,dBm). 
The pump source, an active frequency multiplier (AFM6 70-110+14 from RPG Radiometer Physics), delivered output powers of up to 60\,mW (17.8\,dBm).
The final pump and probe frequencies were both located in the W-band.
We performed some additional measurements with the probe frequency being around 200\,GHz using commercial components by VDI.
It is only the probe transitions that share an energy level with the pump transition that are affected by the additional source.
An additional $1f'$ demodulation of the pumped signal is experimentally realizing a difference spectrum of the on- and off-resonant measurements, or, in other words, a subtraction of a conventional and a DR measurement. In this way, it is only the lines of interest remaining and the assignment of even weak or blended features is facilitated \citep{Zingsheim2021_DMDR}.

\section{Results: Spectroscopic signatures}   
\label{Sec:ExperimentalResults}               

The measurement results for \textit{Aa} and \textit{Ag n}-propanol are presented in this section. 
The central focus of this chapter is the analyses and the resulting quantum mechanical models of rotational spectra, as well as the verification of assignments, since both conformers show perturbed patterns. 
Our assignment procedure benefited, in terms of speed and unambiguity, from the \textit{LLWP} software, which is based on Loomis-Wood plots\footnote{Download at https://llwp.astro.uni-koeln.de/} \citep{LLWP_Luis}.
Adjacent transitions of certain series are plotted around their predicted frequency in Loomis-Wood plots to increase the confidence of correct assignments or to visualize systematic deviations between observed and predicted transition frequencies.

\subsection{\textit{Aa n}-propanol}           

The \textit{Aa} conformer is an asymmetric rotor quite close to the prolate limit with $\kappa=-0.978$ ($\kappa=[2B-A-C]/[A-C]$).
The dipole moments were determined with Stark effect measurements to be $\mu_a=0.21(7)$\,D and $\mu_b=1.54(2)$\,D \citep{Propanol_U5_1970_Abdurakhmanov_123}.
In total, 73 lines (118 transitions) were assigned and fit up to 40\,GHz in a previous study \citep{n-propanol_DreizlerScappini_1981}. Therein, all lines from \citet{Propanol_U2_1968} were remeasured; 
$a$- and $b$-type transitions were assigned and fit and in particular the assignments of weak $a$-type transitions were secured with the help of MW-MW DR measurements \citep{n-propanol_DreizlerScappini_1981}. 
A splitting due to the internal rotation of the methyl group ($-$CH$_3$) could sometimes be resolved (14\,A$+$14\,E transitions) and together with 45 blended doublet features, the transition frequencies were reproduced with a standard deviation of 195\,kHz \citep{n-propanol_DreizlerScappini_1981}.
Thereby, the potential barrier of the methyl group internal rotation was determined to be $V_3=955(21)$\,cm$^{-1}$ (or $2730(60)$\,cal/mol) and the angle of the internal rotation axis $i$ to the principal moment of inertia axis $a$ to be $\sphericalangle(i,a)=(29\pm1)\,^\circ$ \citep{n-propanol_DreizlerScappini_1981}.

\begin{figure}[t]
\centering
\includegraphics[width=0.95\linewidth]{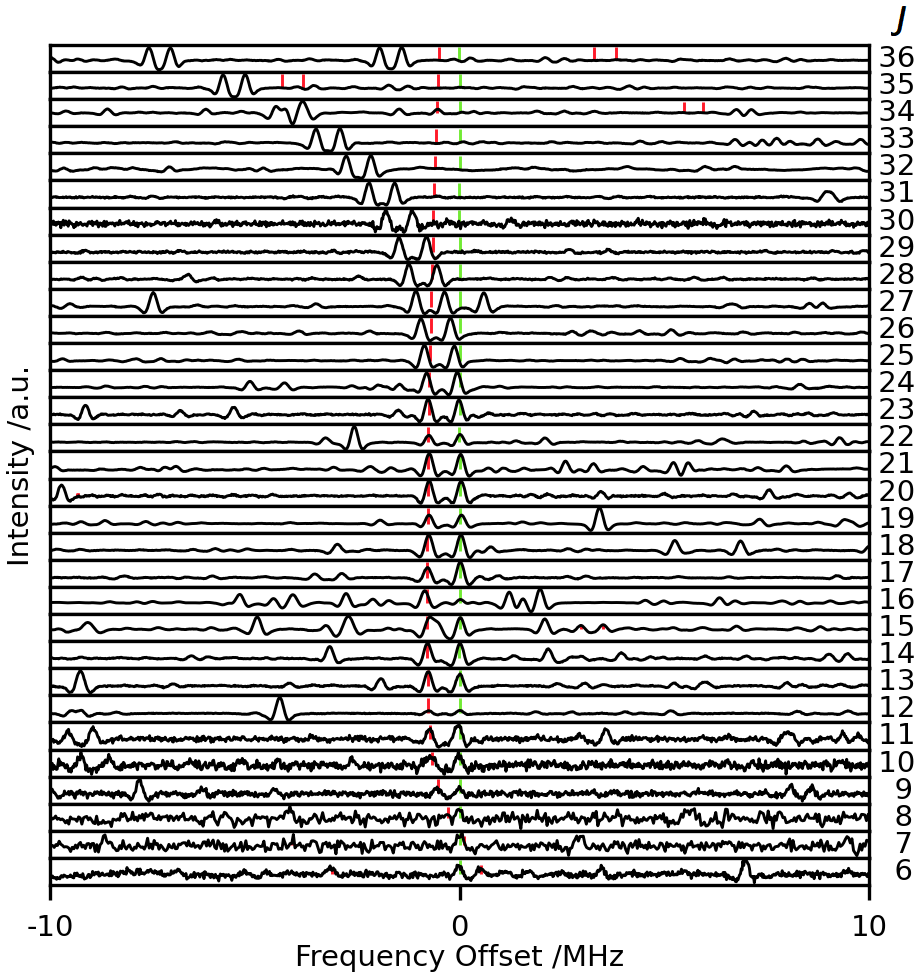}
\caption{Loomis-Wood plot of \textit{Aa n}-propanol - Screenshot of the \textit{LLWP} software \citep{LLWP_Luis}.
Shown here is a doublet series (A and E components of the methyl group internal rotation) of the $J_{K_a,K_c}=J_{3,J-3}\leftarrow J_{2,J-2}$ $b$-type transitions from $J=6$ (bottom panel) to $J=36$ (top panel).
Predictions of the A components are set as center frequencies and are depicted by green sticks. The E components are depicted by red ones.
This $^rQ$ series with $K_a+K_c=J$ and $\Delta K_a,\Delta K_c=+1,-1$ is well reproduced by our spectroscopic model described in Table~\ref{Tab:Spectroscopic_parameters_Aa} up to $J=22$, however, transitions up to $J=41$ ($\Delta\nu\approx 90$\,MHz) are straightforwardly assignable. 
}
\label{Fig2:Aa-series}
\end{figure}

\begin{figure}[t]
\centering
\includegraphics[]{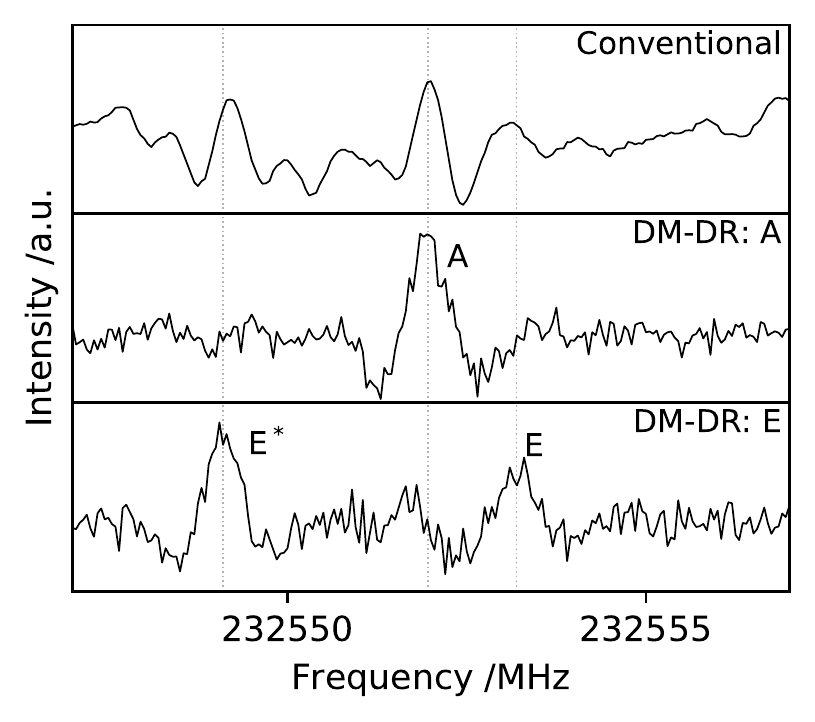} 
\caption{DM-DR measurements of \textit{Aa n}-propanol of the $J_{K_a,K_c}=10_{4,6}\leftarrow9_{3,7}$ transition around 232.55\,GHz.
The lower panels show two DM-DR measurements where the A and E components of the $J_{K_a,K_c}=9_{3,7}\leftarrow9_{2,8}$ transition are used as pump frequencies at 113811.6420\,MHz and 113810.5391\,MHz, respectively. 
A conventional measurement is shown in the upper panel.
E$^*$ transitions, which are forbidden in rigid rotors \citep{Herschbach_PO_forbidden_E_lines}, secure the assignment of internal rotation components.
We note that slightly off-resonant pumping leads to small, asymmetric features of A and E components in the DM-DR spectrum of the other one, as pumped A and E components are rather close in frequency, cf. \citet{Zingsheim2021_DMDR}.
}
\label{Fig3:DM-DR-200}
\end{figure}

We used the available parameters and transition frequencies \citep{n-propanol_DreizlerScappini_1981} to set up an initial spectroscopic model of an asymmetric rotor, including three-fold internal rotation in ERHAM \citep{Groner_1992,Groner_1997,Groner_review}. Thanks to the initial prediction and with the application of an iterative assignment and fit strategy, we could assign strong and also weak $b$-type transitions in a straightforward fashion (see Fig.~\ref{Fig2:Aa-series}).
Transitions from \citet{n-propanol_DreizlerScappini_1981} were remeasured if possible.\footnote{Two $Q$-branch $a$-type lines (or four transitions) were not included in the analysis of \citet{n-propanol_DreizlerScappini_1981}, but were secured by them with MW-MW DR spectroscopy; These lines are successfully incorporated to the fit in this study. However, six other transitions have been unweighted in this study due to conspicuously large deviations.}
Furthermore, we secured assignments of weak $a$-type transitions in the W-band region with the DM-DR technique.
Additionally, transitions of the $^rR$ series ($J_{K_a,K_c}=J_{4,J-3}\leftarrow (J-1)_{3,J-4}$ with $J$ up to 13) have been unambiguously assigned above 200\,GHz with DM-DR measurements (see Fig.~\ref{Fig3:DM-DR-200}).
The assignment of internal rotation components, A and E, is unambiguous as we observed E$^*$ transitions, which are forbidden in rigid rotors \citep{Herschbach_PO_forbidden_E_lines}.
However, the $^rR$ series transitions appeared to be noticeably shifted from their prediction and not all of them are incorporated in the final fit. Including these transitions is possible if others are excluded instead. The final choice of fit transitions is somewhat ambiguous: the quantum number coverage is visualized in Fig.~\ref{FigA_QN-overview} of the Appendix. 
Larger systematic deviations, $\Delta\nu=\nu_{\textrm{Obs}}-\nu_{\textrm{Calc}}$, occur from our final predictions for various series (cf. Fig.~\ref{Fig2:Aa-series}). These deviations start at similar $J$ values for series involving identical $K_a$ values, see Table~\ref{Tab:Aa_QN_overview}.
However, the assignment of these transitions is unambiguous as demonstrated in Fig.~\ref{Fig2:Aa-series} and the assignment of internal rotation components is also secured (cf. Fig.~\ref{Fig3:DM-DR-200}).
Finally, we could assign 1255 transitions with frequencies up to 505\,GHz and $J_{\textrm{max}}=70$, thereof, 928 transitions are well reproduced by the prediction.
The determined spectroscopic parameters can be found in Table~\ref{Tab:Spectroscopic_parameters_Aa}.

\setlength{\tabcolsep}{2pt}
\begin{table}
\footnotesize
\begin{center}
\caption[]{Quantum number overview of the final analysis of \textit{Aa n}-propanol.$^\textrm{a}$}
\label{Tab:Aa_QN_overview}
\begin{tabular}{c c | l c r r r r}
\hline\hline

$K_a$ $^\textrm{b}$ & $K_a''+K_c''$ & $a$-type$^\textrm{c}$ & & \multicolumn{4}{c}{$b$-type$^\textrm{c}$} \\
\cline{3-3}
\cline{5-8}
 & &  \multicolumn{1}{c}{$^qR$} & & \multicolumn{1}{c}{$^rP$} & \multicolumn{1}{c}{$^rQ$} &  \multicolumn{1}{c}{$^rR$} & \multicolumn{1}{c}{$^pR$}  \\
  
      &           & $\pm$0,$-$1 && +1,$-$3 &  +1,$-$1  &    +1,$-$1 &  $-$1,+3   \\ 
      &           & $\pm$0,$-$1 && +1,$-$1 & +1,$-$1   &     +1,+1  &  $-$1,+1    \\
      
\hline
  0   &     $J$    &   16~(16) &&  $-$ & 32~(32)   & 69~(70) &  68~(70)      \\
 & & \\  
  1   &    $J+1$   &   17~(17) &&  7~(7) & 28~(28)   & 20~(20) &  29~(29)      \\
  1   &     $J$    &   16~(16) &&  7~(7) & 31~(41)   & 31~(53) &  41~(53)\\
 & & \\  
  2   &    $J+1$   &   16~(16) && 13~(13) & 20~(26)   & 23~(32) & 26~(35)    \\
  2   &     $J$    &   16~(16) && 13~(13) & 22~(41)   & 20~(37) & 22~(26)\\
 & & \\  
  3   &    $J+1$   &   16~(16) && $-$&~~9~(25)   &  9~(12) &  $-$ \\
  3   &     $J$    &   16~(16) && $-$&~~9~(11)        &  9~(13) &  $-$ \\
 & & \\ 
  4   &    $J+1$   &   11~(11) && $-$& $-$       &  $-$    & $-$     \\
  4   &     $J$    &   11~(11) && $-$& $-$       &  $-$    & $-$     \\
  
\hline
\end{tabular}
\end{center}
\begin{footnotesize}
$^\textrm{a}$ Values represent $J_{\textrm{max}}$, the highest $J$ quantum number of well reproduced transitions of a given series in our final spectroscopic model (Values in parentheses represent $J_{\textrm{max}}$ of assigned transitions); $^qQ$ transitions are neglected here.\\
$^\textrm{b}$ The given $K_a$ is always the lower one of $K_a'$ and $K_a''$.\\
$^\textrm{c}$ Selection rules for $K_a$ and $K_c$ quantum numbers: $\Delta K_a=K_a'-K_a'',\Delta K_c=K_c'-K_c''$. Upper values are for asymmetry side $K_a''+K_c''=J''+1$ and lower ones for $K_a''+K_c''=J''$. 
Additionally, $\Delta K_a$ with $-1,\pm0,+1$ is also represented by the superscripts $p,q,r$, respectively.
The capital letters $P$, $Q$, $R$, represent $\Delta J$ with $-1,\pm0,+1$ respectively.
\end{footnotesize}
\end{table}

\begin{table}
\begin{scriptsize}
\centering
\caption[Spectroscopic parameters of \textit{Aa n}-propanol.$^\textrm{a}$]{Spectroscopic parameters of \textit{Aa n}-propanol.$^\textrm{a}$}
\label{Tab:Spectroscopic_parameters_Aa}
\begin{tabular}{l l S l S S} 
\hline\hline
\multicolumn{2}{l}{Parameter} & \multicolumn{1}{c}{Theory}  & ~~& \multicolumn{2}{c}{Experiment} \\
\cline{3-3}
\cline{5-6}
\multicolumn{2}{l}{(MHz)} & \multicolumn{1}{c}{Ref.~K10} &  & \multicolumn{1}{c}{Ref. DS81}  & \multicolumn{1}{c}{This work}  \\
 \hline
$A$ &           & ~~~~~~26579 &  & 26401.671(50)  & 26401.4395(40)\\
$B$ &       &  3784 &   & 3802.154(11)   &  3802.16025(39) \\
$C$ &           &  3531 &  & 3549.543(20)   &  3549.45427(33) \\
& & & &~\\
$\Delta_{K}$    & x\,$10^{3}$  &   58.7  &  & 141.5(47)    & 149.31(55)\\
$\Delta_{JK}$   & x\,$10^{3}$  &   -2.35 &  &   9.57(99)   & -0.3608(295)\\
$\Delta_{J}$    & x\,$10^{3}$  &    0.828&  &   1.672(73)  & 0.99076(48)  \\
$\delta_K$  & x\,$10^{3}$  &   0.361 &  & -29.6(69)    & -3.250(164) \\
$\delta_J$  & x\,$10^{3}$  & 0.0737&  &   0.0943(45) & 0.129324(170) \\
& & & &~\\
$\Phi_{K} $ & x\,$10^{6}$    & \multicolumn{1}{c}{$-$} &  & \multicolumn{1}{c}{$-$} &   -153.3(202) \\
$\Phi_{KJ}$ & x\,$10^{6}$    & \multicolumn{1}{c}{$-$} &   & \multicolumn{1}{c}{$-$} &   -117.60(210) \\
$\Phi_{JK}$ & x\,$10^{6}$    & \multicolumn{1}{c}{$-$} &   & \multicolumn{1}{c}{$-$} &   4.870(202) \\
$\Phi_{J} $ & x\,$10^{6}$    & \multicolumn{1}{c}{$-$} &   & \multicolumn{1}{c}{$-$} &   -0.00390(48) \\
$\phi_{K}$  & x\,$10^{6}$     & \multicolumn{1}{c}{$-$} &   & \multicolumn{1}{c}{$-$} &   707.1(212) \\
$\phi_{JK}$ & x\,$10^{6}$     & \multicolumn{1}{c}{$-$} &   & \multicolumn{1}{c}{$-$} &   1.850(44) \\
$\phi_{J}$  & x\,$10^{6}$     & \multicolumn{1}{c}{$-$} &   & \multicolumn{1}{c}{$-$} &   -0.002675(230) \\
& & & &~\\
$L_{JK}$   & x\,$10^{9}$    & \multicolumn{1}{c}{$-$} &   & \multicolumn{1}{c}{$-$} &          -119.9(37) \\
$L_{JJK}$  & x\,$10^{9}$     & \multicolumn{1}{c}{$-$} &   & \multicolumn{1}{c}{$-$} &         -1.015(39) \\
$L_J$      & x\,$10^{9}$    & \multicolumn{1}{c}{$-$} &   & \multicolumn{1}{c}{$-$} &         -0.002291(152) \\
$l_{J}$    & x\,$10^{9}$     & \multicolumn{1}{c}{$-$} &  & \multicolumn{1}{c}{$-$} &        -0.001340(85) \\
& & & &~\\              
$\sphericalangle(i,a)$~& /$\,^\circ$ & \multicolumn{1}{c}{$-$} &   & 29(1) & 26.49(38) \\  
$\rho$ & &      \multicolumn{1}{c}{$-$} &  & \multicolumn{1}{c}{$-$} & 0.14924(107) \\
$\epsilon_{10}$ &   &      \multicolumn{1}{c}{$-$} &  & \multicolumn{1}{c}{$-$} & -1.2696(127) \\
\hline

\multicolumn{4}{l}{Number of transitions$^\textrm{b}$}       &   113 & 928    \\
\multicolumn{4}{l}{Number of lines}         &     73 & 689    \\
$rms$  & /\,kHz   &   &  &  195 & 48  \\
$wrms^\textrm{b}$ &    &   &  &   \multicolumn{1}{c}{$-$} & 1.07  \\
\hline
\end{tabular}
\end{scriptsize}
\begin{footnotesize}
\begin{flushleft}
    \vspace{0.5em}
   { References: K10 \citep{n-propanol_Kisiel_2010}; DS81  \citep{n-propanol_DreizlerScappini_1981}\\
    $^\textrm{a}$ Watson's A reduction Hamiltonian of asymmetric rotors is used with ERHAM.
    $A$, $B$, and $C$ are rotational constants; $\Delta$'s, $\Phi$'s, and $L$'s are quartic, sextic, and octic centrifugal distortion constants, respectively.
    The angle $\sphericalangle(i,a)$, the unitless $\rho$ value, and $\epsilon_{10}$ describe splittings due to the internal rotation of the $-$CH$_3$ group.\\
    $^\textrm{b}$ We remeasured 34 transitions of \citet{n-propanol_DreizlerScappini_1981} and unweighted 6 $Q$-branch $a$-type transitions. In total, we assigned 1255 transitions.
    
}
\end{flushleft}
\end{footnotesize}
\end{table}

\subsection{\textit{Ag n}-propanol}           

The three members of the \textit{G} family were extensively described in the literature plus 
the \textit{Aa} conformer to some extent.
The situation is different for the fifth conformer of \textit{n}-propanol: \textit{Ag}.
There is no spectroscopic model of rotational spectra with experimental accuracy available, and only 22 $J_{K_a,K_c}\leftarrow (J-1)_{K_a,K_c-1}$ $^qR$ series transitions (1, 3, 5, and 2 transitions for $J=1$, 2, 3, and 4, respectively, for each of the two tunneling states), were tentatively assigned in the MW region \citep{Viniti_Ag_1976}.
Therein, doublet features were observed and assigned to the tunneling states \textit{Ag$^+$} and \textit{Ag$^-$}. The average $B+C$ was determined to be about 7.3\,GHz \citep{Viniti_Ag_1976}.
A repeating signature with this spacing was also observed in the MMW region
\citep[see Fig. 2 of][]{n-propanol_Kisiel_2010}
and can be attributed to $K_a$ structures with increasing $J$ quantum numbers of the \textit{Ag} conformer. However, to our knowledge no spectroscopic model with experimental accuracy nor assignments are publicly available for that latter work.
Namely, \textit{Ag n}-propanol is a perturbed system that cannot be analyzed easily.
The focus of this work is on providing initial assignments of $^qR$ series transitions, as the dipole moments are calculated to be $\mu_a=1.15$\,D, $\mu_b=0.39$\,D, and $\mu_c=1.15$\,D\footnote{Allowed $a$- and $b$-type transitions occur within a tunneling state: $Ag^+ \leftrightarrow Ag^+$ and $Ag^- \leftrightarrow Ag^-$, whereas $c$-type transitions occur between the tunneling states: $Ag^+ \leftrightarrow Ag^-$.}, but the energy difference of $Ag^+$ and $Ag^-$ tunneling states is unknown so far. Then, \textit{Ag n}-propanol is expected to be close to the prolate limit with $\kappa=-0.98$ \citep{n-propanol_Kisiel_2010}.

\begin{sidewaysfigure*}
\centering
\includegraphics[]{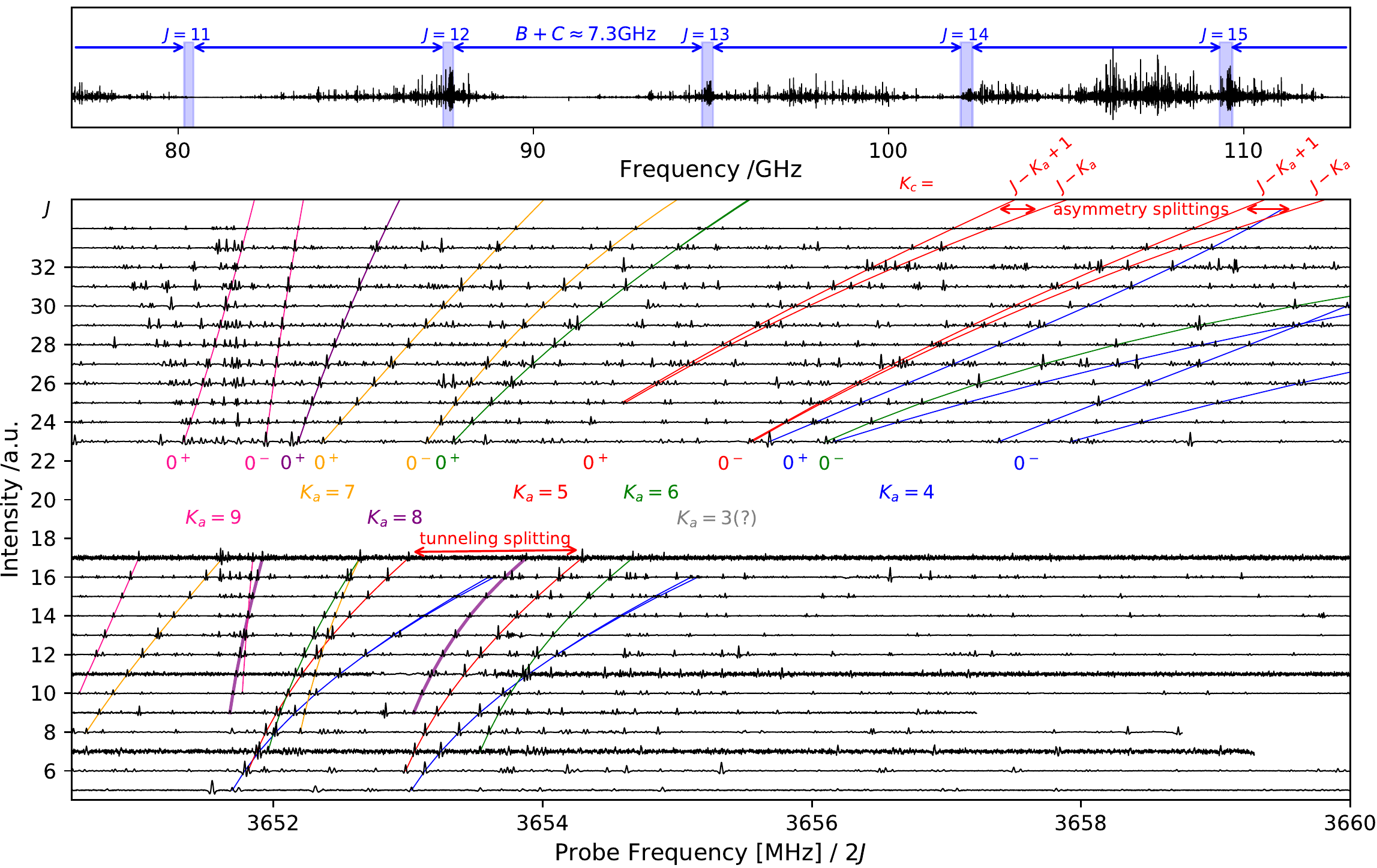} 
\caption{Fortrat diagram of \textit{n}-propanol. \textbf{Top panel:} A broadband measurement in the W-band region is shown, where respective regions with signatures of the \textit{Ag} conformer, leading to the Fortrat diagram, are highlighted. In some regions, the spectrometer is not as sensitive, e.g. around $J=11$, compared to others.
\textbf{Bottom panel:} The colored lines in the Fortrat diagram mark linkages of $^qR_{K_a}$ series transitions, $J_{K_a,K_c}\leftarrow (J-1)_{K_a,K_c-1}$. DM-DR spectroscopy secured assignments in the W-band region (from $J=10$ to $J=16$). The existence of two tunneling states is proven as for each $K_a$ partner series are found; the tunneling states are marked by $0^+$ and $0^-$. More series can be found, but are not assigned yet, e.g. $^qR_{3}$ ones. We note the larger tunneling splitting of $^qR_{6}$ series (green lines), e.g., in comparison to $^qR_{5}$ ones, and the different slopes for $^qR_{8}$ series (thick purple lines).}
\label{Fig4:Ag-Fortrat}
\end{sidewaysfigure*}

We performed CP-FTMW spectroscopy in particular to check the assignments of $3_{K_a,K_c}\leftarrow 2_{K_a,K_c-1}$ $a$-type transitions from \citet{Viniti_Ag_1976}.
We measured not only the aforementioned transitions of \textit{Ag n}-propanol, but also of  gauche-propanal as its resulting tunneling spectroscopic pattern is already well-understood \citep{ZINGSHEIM2021_Propanal}. Similar spectroscopic signatures are expected for both molecules and by comparing both spectra; literature assignments of \textit{Ag n}-propanol could not be confirmed, but the first tentative assignments of typical doublet transitions, originating in $Ag^+$ and $Ag^-$ tunneling states, could be made (cf. Fig.~\ref{FigA:Propanol_vs_Propanal} in the appendix).

DM-DR measurements secured the linkages of various transitions that occur about every 7.3\,GHz in the W-band region. 
Afterwards, we assigned $^qR$ series transitions in the frequency region from 36 to 505\,GHz with the help of the \textit{LLWP} software \citep{LLWP_Luis}.
Some of the $^qR$ series of the \textit{Ag} conformer are visualized exemplarily in a Fortrat diagram (Fig.~\ref{Fig4:Ag-Fortrat}).
Assignments of $^qR_{K_a}$ series transitions, $J_{K_a,K_c}\leftarrow J-1_{K_a,K_c-1}$, are done as they start in rows of the Fortrat diagram with $J=K_a+1$ ($K''_a\leq J''$).
Two patterns are found, with each consisting of a doublet series for each $^qR_4$, $^qR_5$, $^qR_6$, $^qR_7$, $^qR_8$, and $^qR_9$ series ($4\leq K_a\leq9$) proving the existence of $0^+$ and $0^-$ tunneling states (see Fig.~\ref{Fig4:Ag-Fortrat}).
The nomenclature $0^+$ and $0^-$ is used to highlight that the vibrational ground state $\upsilon=0$ is observed. We assigned the tunneling state to transitions with $4\leq K_a \leq 9$ in the way that we assumed $0^+$ transitions to be always lower in frequency than ones of $0^-$,  which is usually the case if identical transitions are compared (transitions with same $J$'s, $K_a$'s, and $K_c$'s).

\begin{figure}[t]
    \begin{subfigure}[h]{\linewidth}
    \centering
    \includegraphics[width=0.8\linewidth]{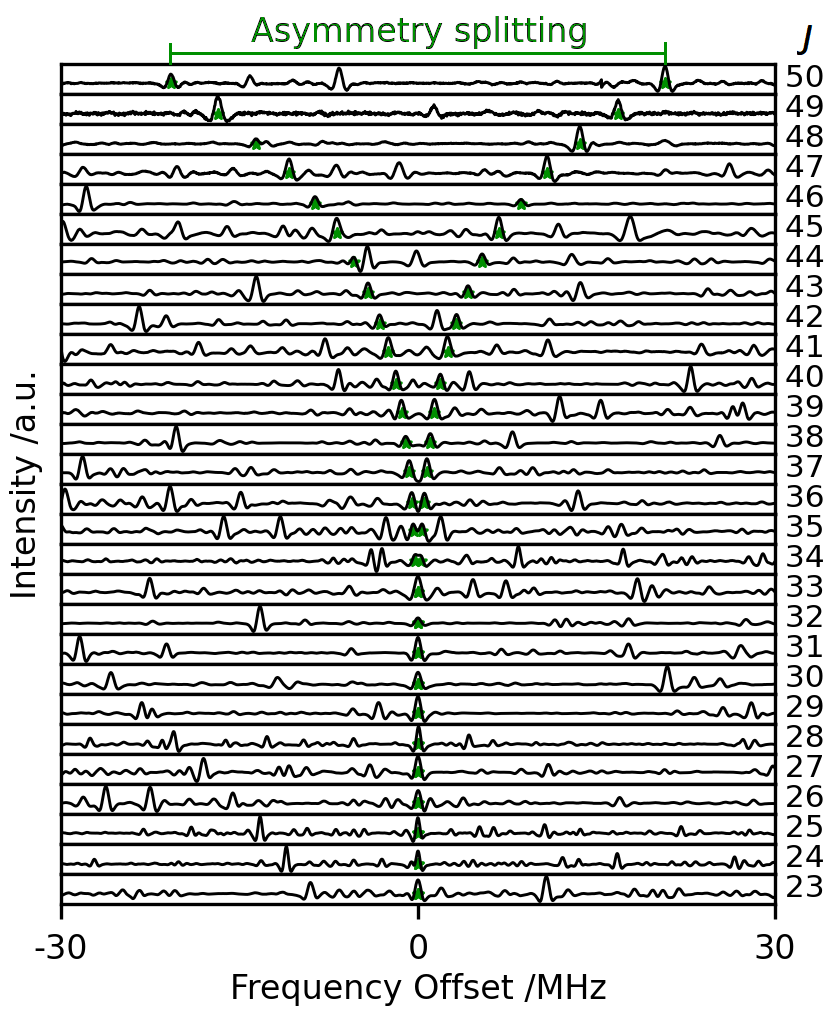}\\
    \caption{Loomis-Wood plot of $^qR_6$ series transitions.}
    \end{subfigure}
    \begin{subfigure}[h]{\linewidth}
    \centering
    \includegraphics[width=0.8\linewidth]{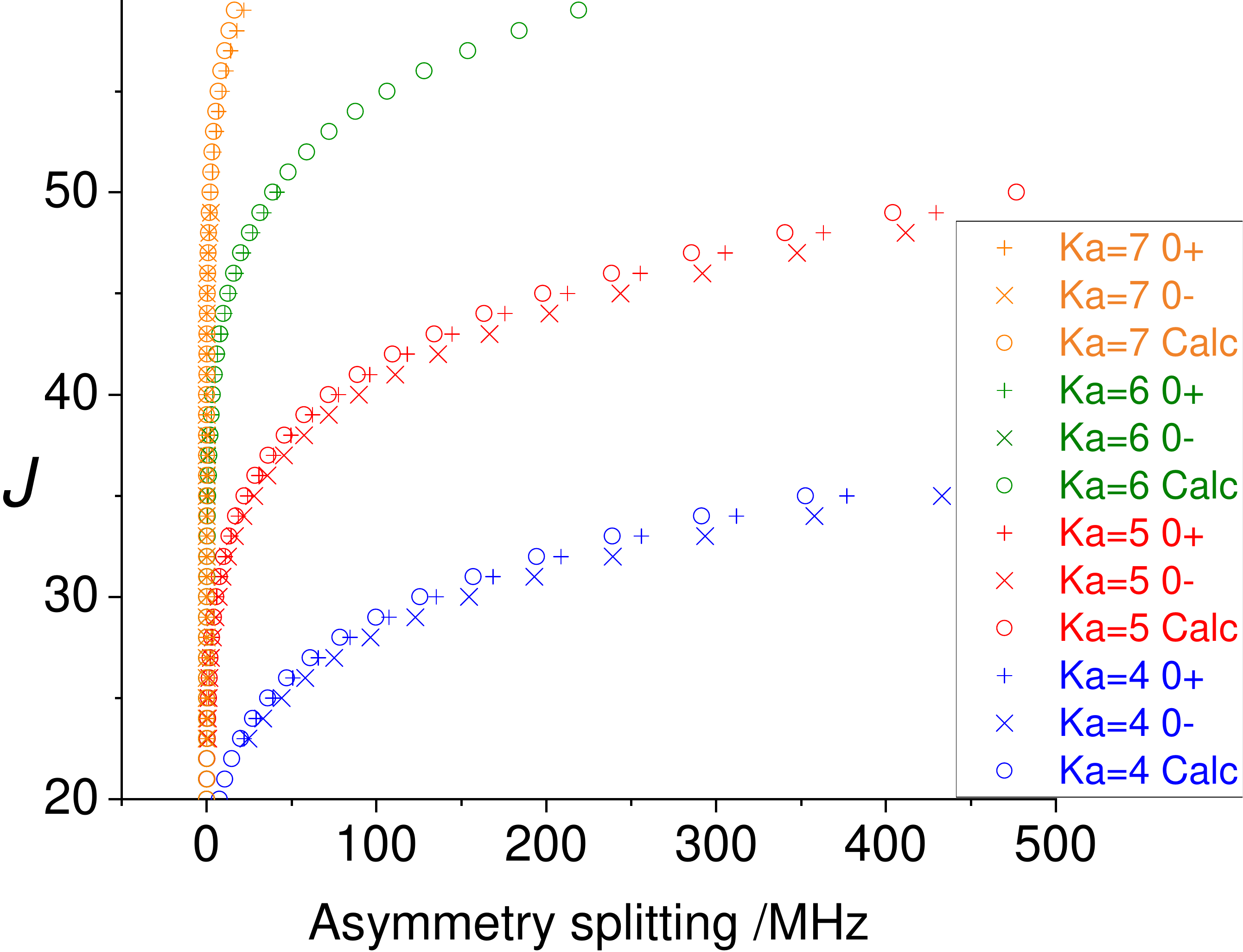}
    \caption{Observed and calculated asymmetry splitting.}
    \end{subfigure}
\caption{Asymmetry splitting of the $^qR_4$, $^qR_5$, $^qR_6$, and $^qR_7$ series transitions.
\textbf{(a)} Exemplary Loomis-Wood plot for $^qR_6$ transitions of $0^+$. Two transitions, either with $K_c=J-K_a+1$ or with $K_c=J-K_a$, are observed and marked by green stars in each row. The average frequency of both transitions is the center frequency (Frequency Offset is 0\,MHz). Screenshot of the \textit{LLWP} software \citep{LLWP_Luis}. The frequency difference of both components is the asymmetry splitting for $K_a=6$ of $0^+$ for a given $J$.
\textbf{(b)} Observed and calculated asymmetry splittings for $^qR_4$ (blue), $^qR_5$ (red), $^qR_6$ (green) and $^qR_6$ (orange). Similar observed and calculated asymmetry splittings confirm the correct assignment of $K_a$ quantum numbers. 
We note that small asymmetry splittings are not resolvable, both transitions are partly blended, and missing experimental values for $J=17-22$ originate in the experimental gap from 130$-$170\,GHz.
}
\label{Fig5:Ag-Asym_splitting}
\end{figure}

A further confirmation of unambiguous assignments is the observed asymmetry splitting for $^qR_4$, $^qR_5$, $^qR_6$, and $^qR_7$ series.
The two asymmetry components, transitions with identical $J$ and $K_a$ but either $K_c=J-K_a$ or $K_c=J-K_a+1$, are frequently blended for low $J$ and higher $K_a$ values (prolate pairing), but this degeneracy is lifted for higher $J$ (cf. colored lines in Fig.~\ref{Fig4:Ag-Fortrat} which are split into two lines at higher $J$).
The asymmetry splitting is the frequency difference of the two $a$-type transitions with $J_{K_a,J-K_a}\leftarrow J-1_{K_a,J-K_a-1}$ and $J_{K_a,J-K_a+1}\leftarrow J-1_{K_a,J-K_a}$.  
The predicted asymmetry splittings, from the ab initio calculations of \citet{n-propanol_Kisiel_2010}, show the right order of magnitude when compared to the observed ones of both tunneling states; more importantly, they show similar systematic deviations for all $K_a$ numbers (see Fig.~\ref{Fig5:Ag-Asym_splitting}).

At this point, we tried to reproduce the determined frequencies of assigned transitions with $4\leq K_a\leq9$ by applying a common asymmetric rotor Hamiltonian, which is comparable to what has been done for the \textit{Aa} conformer (cf. Table~\ref{Tab:Spectroscopic_parameters_Aa}); This time the \mbox{\textit{SPFIT}} program was used \citep{PICKETT1991}.
Many series seem to deviate from asymmetric rotor spectra.
We expect identical $^qR_{K_a}$ series (same $K_a$ and same asymmetry side) of the two tunneling states to appear as nearly parallel lines in Fig.~\ref{Fig4:Ag-Fortrat}, whose spacing should decrease for increasing $K_a$ quantum numbers. This is not the case, for instance, the two series with $K_a=8$ (purple lines) have quite different slopes and the tunneling splitting of series with $K_a=6$ (green lines) is larger than that of the series with $K_a=5$ (red lines).
We tried to incorporate even order Coriolis-type parameters ($F_{bc}$, $F_{ac}$, and $F_{ab}$) to account for possible tunneling-rotation interaction between the two tunneling states, $Ag^+$ and $Ag^-$, as  has been done successfully several times for similar molecular systems in the past; for instance, in \citet{RAS_Coriolis,B304566H,ethanol_gauche_1997,ZINGSHEIM2021_Propanal}. However, for the case of \textit{Ag n}-propanol, still no experimental accuracy for our predictions has been reached.

We continued assigning $J$, $K_a$, and $K_c$ quantum numbers to transitions with $K_a<3$.
However, transitions from one of the two series, $Ag^+$ and $Ag^-$, are not strictly lower in frequency than the other, in contrast to series with $4\leq K_a \leq 9$. Crossing series are observed in Loomis-Wood plots (Figs.~\ref{FigA:Fortrat_broad}-\ref{FigA:LW_Ka2} in the appendix).
Therefore, the tunneling state assignments remain speculative for transitions with $K_a<3$.
The two tunneling series ($^qR$ series with identical $K_a$ and $K_c$ quantum numbers) with $K_a<3$ are even less parallel than observed ones with $4\leq K_a \leq 9$.
We expect states with $K_a<3$ to be strongly influenced by tunneling-rotation interaction between $Ag^+$ and $Ag^-$ tunneling states, cf. Fig.~\ref{FigA_red-egy}. 
Finally, we could not securely assign $^qR_3$ transitions as too many candidate series are close in frequency (Fig.~\ref{FigA:LW_Ka3}).

In summary, we could prove the existence of the two tunneling states $Ag^+$ and $Ag^-$ and assign lines with $J_\textrm{max}=69$ and $K_{a,\textrm{max}}=9$. A quantum number overview of assigned transitions is visualized in Fig.~\ref{FigA_QN-overview}. 
Deriving a spectroscopic quantum mechanical model with experimental accuracy is beyond the scope of this work because of the multiple perturbations whose origins are unknown at present.


\section{Discussion}                      
\label{Sec:Discussion}                    

In this work, we present an extended spectroscopic quantum mechanical model based on additional assignments in the MMW region for \textit{Aa n}-propanol and secured assignments of both tunneling states, $Ag^+$ and $Ag^-$, for \textit{Ag n}-propanol.
In the following, we discuss the spectroscopic results of the \textit{Aa} and \textit{Ag} conformers (Sects.~\ref{Sec:Dis_Aa} and \ref{Sec:Dis_Ag}). Subsequently, we offer some ideas to overcome certain limitations, specifically, thanks to a global analysis of both conformers (Sect.~\ref{Sec:Dis_global}).
Next, spectroscopic results of \textit{n}-propanol are judged in regard of astronomical searches (Sect.~\ref{Sec:Disc_astro_search}). The rotational data are used in a companion article for its astronomical detection (Sect.~\ref{Sec:Dis_astro_det}).
Moreover, we showed that DM-DR measurements can straightforwardly be extended to frequencies higher than the W-band, see Fig.~\ref{Fig3:DM-DR-200}, as was demonstrated in the original DM-DR manuscript \citep{Zingsheim2021_DMDR}.

\subsection{\textit{Aa n}-propanol}       
\label{Sec:Dis_Aa}                        

We extended spectroscopic assignments of \textit{Aa n}-propanol up to 505\,GHz with $J_{\textrm{max}}=70$ and ${K_a}_{\textrm{,max}}=6$. Finally, 928 transitions are well reproduced by the spectroscopic parameters in Table~\ref{Tab:Spectroscopic_parameters_Aa}.
Larger systematic deviations $\Delta\nu$ from our predictions occur for certain series (cf. also Fig.~\ref{Fig2:Aa-series}); deviations start from transitions with similar $J$ when series with identical $K_a$ values are compared.
The deviations occur at rather low $K_a$ values and the origins of these perturbations are unknown so far.
Transitions involving $K_a=0$ and 1 seem to be unperturbed as $b$-type transitions involving  $K_a=1\leftrightarrow 0$ are well reproduced in our final spectroscopic model.
On the other hand, $b$-type transitions with $K_a=2\leftrightarrow 1$, $K_a=3\leftrightarrow 2$, and $K_a=4\leftrightarrow 3$ are only well reproduced up to around $J=30$, $J=20$, and $J=10$, respectively (see Table~\ref{Tab:Aa_QN_overview}). This case is similar to that of \textit{a}-ethanol, as discussed for the $^{13}$C isotopomers \citep{BOUCHEZ}.

If different states are interacting, consecutive transitions of one series are frequently perturbed. In particular, systematic deviations occur around the strongest perturbed energy levels. So-called avoided crossing patterns may be observed; in this case, transitions belonging to the same series with somewhat smaller and larger quantum numbers than heavily perturbed ones can often be assigned with confidence and, more importantly, reproduced to experimental accuracy without treating the interaction.
In this way, perturbed energy levels can be located and transitions from these levels can be unweighted in the analysis if the origin of the interaction is unknown or indescribable.
For instance, at the start of the analysis of \textit{}gauche-propanal, certain $\upsilon=0$ series had to be unweighted \citep{ZINGSHEIM2017_Propanal}, but they could be properly described by including Fermi resonances and Coriolis interaction in a global analysis with $\upsilon_{24}=1$ \citep{ZINGSHEIM2021_Propanal}.
In the case of \textit{Aa n}-propanol, no transitions with higher $J$ values than the ones that deviate from the prediction could be fit to experimental accuracy; the strongest perturbed energy levels and therefore the center of the perturbation, or the strongest perturbed levels and their exact energies, are unknown to this point. 
Therefore, we expect that our final spectroscopic model is effective since already perturbed transitions may be included in the line list. 

\subsection{\textit{Ag n}-propanol}       
\label{Sec:Dis_Ag}                        

We thoroughly examined assignments of \textit{Ag n}-propanol in the MW and MMW regions and proved the existence of the two tunneling states \textit{Ag}$^+$ and \textit{Ag}$^-$.
First, DM-DR measurements linked various $^qR$ series transitions and assignment of these series was undertaken with the support of the \textit{LLWP} software.
We could not confirm doublet patterns from the literature \citep[cf. Fig.~\ref{FigA:Propanol_vs_Propanal};][]{Viniti_Ag_1976}, but we did find others instead (see Fig.~\ref{Fig4:Ag-Fortrat}).
The ab initio calculations \citep{n-propanol_Kisiel_2010} allowed us to pinpoint patterns and, in particular, the verification of assignments by studying the asymmetry splitting, cf. Fig.~\ref{Fig5:Ag-Asym_splitting}.
In this way, assignments up to $J_{\textrm{max}}=69$ and ${K_a}_{\textrm{,max}}=9$ with frequencies up to 505\,GHz have been achieved.
A derivation of a spectroscopic quantum mechanical model to fit the experimental data accurately is beyond the scope of this work.

\subsection{Global analysis: \textit{Aa},          
\textit{Ag}$^+$, and \textit{Ag}$^-$ conformers}   
\label{Sec:Dis_global}                             

In the future, \textit{Aa n}-propanol should be described in a more sophisticated spectroscopic quantum mechanical model and an initial description of \textit{Ag n}-propanol with experimental accuracy should be derived. This may be possible only through a global analysis of \textit{Aa}, \textit{Ag}$^+$, and \textit{Ag}$^-$ conformers.
The rotational energy levels of the \textit{G} family seem to be isolated from \textit{A} conformers ones \citep{n-propanol_Kisiel_2010}, therefore, an additional incorporation of \textit{G} conformers in a global analysis is expected to be unnecessary.
A global analysis of the \textit{A} family is expected to be successful only if Coriolis interaction and Fermi resonance between the \textit{Aa} conformer and \textit{Ag}$^+$ or \textit{Ag}$^-$ tunneling states and, very importantly, also tunneling-rotation interactions between the \textit{Ag}$^+$ and \textit{Ag}$^-$ tunneling states are considered.\footnote{Future analyses may also benefit by taking into account preliminary results of Kisiel et. al., presented at the International Symposium on Molecular Spectroscopy (ISMS) 2006. We were made aware of this after submission; assignments are not publicly available.}
The first interaction terms which should be investigated are the ones used to describe the system anti, gauche\textit{}$^+$, and gauche\textit{}$^-$ ethanol \citep{PEARSON2008394}, as the group theoretical considerations and probably also the energy splittings are similar.\footnote{Caution is advised concerning the number of interaction parameters; Using too many leads to high correlations and convergence problems. Using too few may lead to a more stable fit, but may not account for all perturbations, or, if it does, at the expense of unphysical parameters.}
A quantum mechanical treatment taking interactions into account would also allow to determine relative energies as has been done for the \textit{G} family \citep{n-propanol_Kisiel_2010}.
Furthermore, the potential of the \mbox{-OH} group rotation may be studied in detail by taking also higher vibrationally excited states into account.

\subsection{Astronomical search of \textit{n}-propanol}  
\label{Sec:Disc_astro_search}                            

The partition function used for creating the predictions of all five conformers, given in the supplementary material, is $Q_\textrm{Gx}=513984.0861$ ($log~Q=5.7109$) at 300\,K.\footnote{Our predictions (\textit{*.cat} files) will be available together with further information; cf. Appendix~\ref{Sec:A_partition_function}, in the CDMS soon.} This value is derived from the extensive parameter set from \citet{n-propanol_Kisiel_2010} and takes the three \textit{G} conformers and their relative energies into account. Thereby, the degeneracy factor was set to 4 (2 for the internal rotation of the $-$CH$_3$ group plus 2 when taking the mirror images into account).
Conformational correction factors taking into account the existence of the \textit{A} conformers as well as vibrational correction factors, can be found in Table~\ref{Tab:Correction_factors} in the appendix.

Predictions of \textit{Ga}, \textit{Gg}, and \textit{Gg'} are based on the parameter set of \citet{n-propanol_Kisiel_2010}. We note that the relative energy of \textit{Ga} was set to $E_{Ga}=0$\,cm$^{-1}$ (consequently, $E_{Gg}=47.8$\,cm$^{-1}$ and $E_{Gg'}=50.8$\,cm$^{-1}$), instead of using \textit{Gg} as reference point, as done in the original data. 
This adaption is inevitable to derive meaningful intensities and lower state energies in the predictions. Predictions in Pickett's SPCAT format \citep[\textit{*.cat} files;][]{PICKETT1991} can be found in the supplementary material and will be added to the CDMS as well.\\
The predictions of \textit{Aa n}-propanol are limited to transitions with $J$ up to 70, 30, 20, and 10 for $K_a=1$, 2, 3, and 4, respectively. All assigned transitions with $K_a\leq5$ can be used in astronomical observations, but measured transition frequencies should replace predicted ones. Transitions with $K_a>5$ should not be used in astronomical observations due to the limitations of our spectroscopic quantum mechanical model.
The intensities and lower state energies in our predictions were corrected by an assumed relative energy of $E_{Aa}=30$\,cm$^{-1}$.
Overall, the number of unambiguously assigned transitions up to 505\,GHz and the quantum number coverage for low $K_a$ values is sufficient to search for rotational transitions of the \textit{Aa} conformer in space. \\
An astronomical detection of \textit{Ag n}-propanol is facilitated thanks to many unambiguously assigned transitions. For creating meaningful predictions, despite the fact that a quantum mechanical model with experimental accuracy is missing, we used, in a first step, ab initio calculations that are based on \citet{n-propanol_Kisiel_2010} to derive theoretical predictions.
Even though predicted frequencies do not match experimental accuracy at all, meaningful physical parameters, such as the intensities, lower-state energy, and degeneracies of transitions are obtained by assuming a relative energy of $E_{Ag}=45.5$\,cm$^{-1}$.
Finally, we replaced the predicted frequencies by the measured ones and neglected all unassigned transitions.
In doing so, an astronomical detection of \textit{Ag n}-propanol is possible, but deriving a proper spectroscopic quantum mechanical model should be done in particular to derive relative energies and to enlarge the number of usable transitions.

\subsection{Astronomical detection of \textit{n}-propanol}  
\label{Sec:Dis_astro_det}                                   

Some members of our team searched for distinctive rotational signatures of all five conformers of \textit{n}-propanol in a companion astronomical study that uses the survey called Re-exploring Molecular Complexity with ALMA \citep[ReMoCa; details given in][]{Belloche_2019} toward Sagittarius (Sgr) B2(N) \citep{Belloche_2022}.
This astronomical study reports the identification of \textit{n}-propanol toward the secondary hot molecular core, Sgr B2(N2), thanks to the spectroscopic predictions from this work. The detection relies on eight and five lines of the \textit{Gg'} and \textit{Ag} conformers, respectively. The catalogs produced in this work were instrumental in securing this first astronomical detection of \textit{n}-propanol in a hot core.

\section{Conclusion}                          
\label{Sec:Conclusion}                        

We recorded spectra of \textit{n}-propanol in the region of 18$-$505\,GHz, which allowed us to analyze the rotational signatures of its \textit{Aa} and \textit{Ag} conformers, in addition to the already extensively studied  \textit{Ga}, \textit{Gg}, and  \textit{Gg'} conformers \citep{n-propanol_Kisiel_2010}.
Therefore, rotational signatures of all five conformers of \textit{n}-propanol are available at present and the main conclusions of this work are as follows:
\begin{enumerate}
\item An extended quantum mechanical model of rotational spectra of \textit{Aa n}-propanol has been made available.
\item The existence of the two tunneling states, \textit{Ag}$^+$ and \textit{Ag}$^-$, is proven for the first time by unambiguously assigning transitions of \textit{Ag n}-propanol.
\item A global analysis of \textit{Aa} and \textit{Ag} conformers is recommended if the spectroscopic description of the \textit{Aa} conformer needs to be extended or if a quantum mechanical model with experimental accuracy of the \textit{Ag} conformer is the target.
\item Predictions of the rotational spectra of all five conformers of \textit{n}-propanol (\textit{*.cat} files in the supplementary material) enable their astronomical search.
\end{enumerate}

In addition, the spectroscopic analysis presented in this work allowed for the astronomical detection of \textit{n}-propanol toward a hot core in a companion article \citep{Belloche_2022}.

\begin{acknowledgements}
We would like to thank Dr. Matthias Justen and Dr. Leonid Surin for their help in translating Russian manuscripts and providing literature data from the VINITI database.
B.~H. acknowledges financial support by the Deut\-sche For\-schungs\-ge\-mein\-schaft (DFG; project ID WE 5874/1-1).
This work has been supported via Collaborative Research Centre 956, sub-projects B3 and B4, and the ``Cologne Center for Terahertz Spectroscopy", both funded by the DFG (project IDs 184018867 and SCHL 341/15-1, respectively). 
\end{acknowledgements}

\bibliographystyle{aa} 
\bibliography{main_aa.bib} 

%

\begin{appendix} 

\section{Partition function $Q$ of \textit{n}-propanol}  
\label{Sec:A_partition_function}  

We calculated the partition function of the three \textit{G} conformers $Q_\textrm{Gx}$ with SPCAT to be 513984.0861 at 300\,K, which takes into account the \textit{Ga}, \textit{Gg}, and \textit{Gg'} conformers and a degeneracy factor of 4 (a factor of 2 for the internal rotation of the $-$CH$_3$ group and another factor of 2 for mirror images of the respective conformers; see next paragraph for more details).\footnote{Reducing the total degeneracy factor of 4 is not possible if only one column density for all five conformers of \textit{n}-propanol are to be derived, e.g., in local thermal equilibrium conditions in space, as the \textit{Aa} conformer does not exist in a mirror image form and A and E internal rotation components are resolved.} This value is based on the quantum mechanical description from \citet{n-propanol_Kisiel_2010}.
The spectroscopic quantum mechanical models of \textit{Aa} and \textit{Ag} conformers are not as accurate as the ones of the \textit{G} family and the relative energies of $E_\textrm{Aa}$ and $E_\textrm{Ag}$ to $E_\textrm{Ga}$ are somewhat uncertain, thus, we used the partition function derived from the \textit{G} family $Q_\textrm{Gx}$ for creating predictions (\textit{*.cat} files) of all five conformers. 
For the predictions of \textit{Aa} and \textit{Ag} conformers, the degeneracy factor is 2 and 4, respectively. Both conformers have an internal rotor (A and E transition, whereby both components are blended for the \textit{Ag} conformer), but the \textit{Aa} conformer does not occur in a mirror configuration as the other four conformers do. In this way, we derived correct relative intensities of all five conformers.

\begin{table*}
\centering
\caption[]{Partition function $Q_\textrm{Gx}$ of all three \textit{G} conformers (with spin weight 4), $Q_\textrm{Ga~(1)}$ of \textit{Ga} (with spin weight 1), together with conformational correction factors $F_{Conf}$, vibrational correction factors $F_{Vib}$, and their product $F_{Tot}$ for various temperatures of interest for an astronomical detection.}
\label{Tab:Correction_factors}
\begin{tabular}{l S S S S S S}
\hline\hline
$T$   &   $Q_\textrm{Gx}$ & log~$Q_\textrm{Gx}$ & $Q_\textrm{Ga~(1)}$ & $F_{Conf}$ & $F_{Vib}$ & $F_{Tot}$  \\
(K) \\
\hline
300   &     513984.0861  &  5.7109  & 49318.3481 & 1.4748 &   6.4531 &    9.5167 \\
225   &     318241.7669  &  5.5028  & 32031.6352 & 1.4671 &   3.2754 &    4.8055 \\
200   &     260495.0562  &  5.4158  & 26839.0622 & 1.4631 &   2.6370 &    3.8583 \\
180   &     217323.4156  &  5.3371  & 22911.7137 & 1.4590 &   2.2291 &    3.2523 \\
160   &     177034.0151  &  5.2481  & 19197.9679 & 1.4537 &   1.8966 &    2.7571 \\
150   &     158035.3077  &  5.1988  & 17425.1609 & 1.4504 &   1.7549 &    2.5454 \\
140   &     139840.8805  &  5.1456  & 15710.7948 & 1.4466 &   1.6280 &    2.3551 \\
120   &     106007.9447  &  5.0253  & 12465.7151 & 1.4367 &   1.4146 &    2.0325 \\
100   &      75871.4993  &  4.8801  &  9481.9699 & 1.4221 &   1.2499 &    1.7775 \\
75    &      44088.0429  &  4.6443  &  6158.4144 & 1.3905 &   1.1058 &    1.5376 \\
37.5  &      11387.5229  &  4.0564  &  2178.3969 & 1.2546 &   1.0067 &    1.2630 \\
18.75 &       3228.9908  &  3.5091  &   771.4138 & 1.0769 &   1.0000 &    1.0770 \\
9.375 &       1095.9636  &  3.0398  &   273.6960 & 1.0059 &   1.0000 &    1.0059 \\
\hline
\end{tabular}
\begin{flushleft}
    \vspace{0.5em}
    {
}
\end{flushleft}
\end{table*}

For predicting correct absolute intensities, which are fundamental for deriving meaningful column densities of molecules in space, the existence of both \textit{A} conformers and all vibrational states need to be taken into account in order to derive $Q_\textrm{Tot}$. Therefore, we present conformational and vibrational correction factors in Table~\ref{Tab:Correction_factors}.
The conformational correction factors are derived as follows: 
\begin{align}
F_\textrm{Conf}^T 
&=  \frac{Q_\textrm{Ga~(1)}^T \cdot f^\textrm{'}_\textrm{Aa} \cdot f^\textrm{''}_\textrm{Aa} \cdot BF_\textrm{Aa}^T
+ Q_\textrm{Ga~(1)}^T \cdot f^\textrm{'}_\textrm{Ag} \cdot f^\textrm{''}_\textrm{Ag} \cdot BF_\textrm{Ag}^T 
+ Q_\textrm{Gx}^T }{Q_\textrm{Gx}^T} \nonumber \\
&=  \frac{Q_\textrm{Ga~(1)}^T \cdot 2 \cdot 1 \cdot BF_\textrm{Ag}^T+ Q_\textrm{Ga~(1)}^T \cdot 2 \cdot 2 \cdot BF_\textrm{Ag}^T + Q_\textrm{Gx}^T }{Q_\textrm{Gx}^T},
\end{align}
where the partition function $Q_\textrm{Ga~(1)}^T$ is calculated with degeneracy factor 1, for instance, to be $Q_\textrm{Ga~(1)}=49318.3481$ at 300\,K. The first factor $f^\textrm{'}=2$ takes the internal rotation into account and the second factor , namely, $f^\textrm{''}=1$ or 2, the conformer multiplicity of \textit{Aa} and \textit{Ag}, respectively. 
The Boltzmann factor $BF_\textrm{conf}^T=exp(-E_\textrm{conf}/k_BT)$ of both conformers is derived by assuming $E_\textrm{Aa}=30.0\,$cm\,$^{-1}$ and $E_\textrm{Ag}=45.5\,$cm\,$^{-1}$.
The vibrational correction factors were calculated using fundamental vibrational frequencies (with $\nu\leq1200$\,cm$^{-1}$) of the \textit{Aa} conformer \citep{1968JMoSp..26..368F}, that is, 
$\nu_{30}=   131$\,cm$^{-1}$,
$\nu_{29}=   240$\,cm$^{-1}$,
$\nu_{28}=   286$\,cm$^{-1}$,
$\nu_{27}=   332$\,cm$^{-1}$,
$\nu_{26}=   463$\,cm$^{-1}$,
$\nu_{25}=   758$\,cm$^{-1}$,
$\nu_{24}=   858$\,cm$^{-1}$,
$\nu_{23}=   898$\,cm$^{-1}$,
$\nu_{22}=   971$\,cm$^{-1}$,
$\nu_{21}=  1013$\,cm$^{-1}$,
$\nu_{20}=  1047$\,cm$^{-1}$, and $\nu_{19}=  1066$\,cm$^{-1}$.
Applying the correction factors from Table~\ref{Tab:Correction_factors} allows to derive accurate column densities of \textit{n}-propanol and derive relative abundances to other molecular species in astronomical sources.

\section{Rotational signatures of \textit{Ag~n}-propanol}  

\begin{figure*}[t]
\centering
\includegraphics[width=1.0\linewidth]{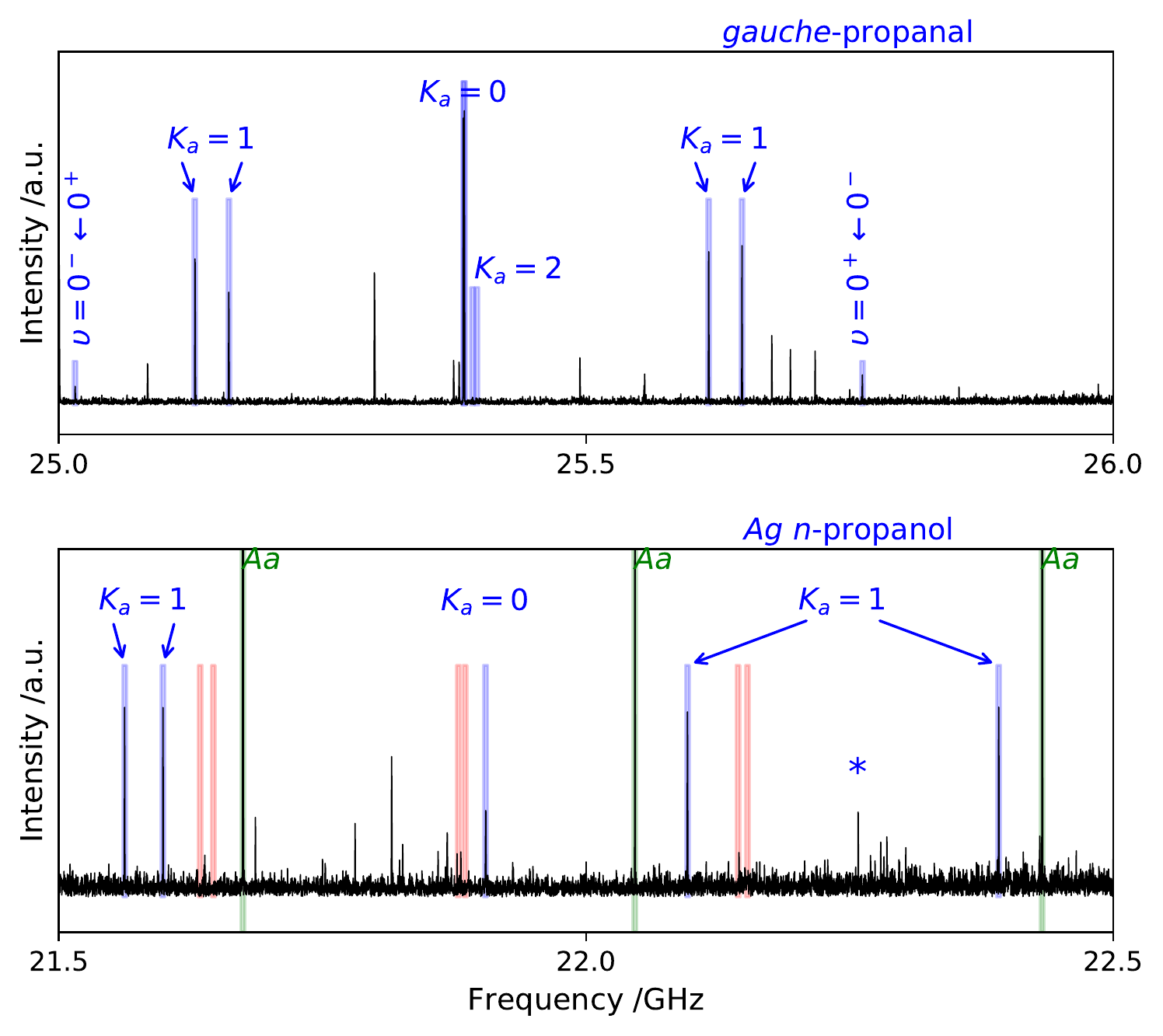} 
\caption{Microwave spectra of \textit{n}-propanol and propanal. 
Assignments of $a$-type transitions with $3_{K_a,K_c}\leftarrow 2_{K_a,K_c-1}$ are shown for both molecules ($K_a$'s are given in the figure).
\textbf{Top panel:} Rotational spectrum of gauche-propanal with assignments from \citet{ZINGSHEIM2021_Propanal} marked in blue.
\textbf{Bottom panel:} Assigned doublet transitions for \textit{Ag n}-propanol from \citet{Viniti_Ag_1976} are marked in red, but cannot be verified in our CP-FTMW jet experiment. Our tentative assignments of $K_a=1$ transitions of \textit{Ag n}-propanol are marked in blue. We note that the transition marked with an asterisk (*) is another possible candidate for a transition with $K_a=1$, more information can be found in the text.}
\label{FigA:Propanol_vs_Propanal}
\end{figure*}

\begin{figure*}[h]
\centering
\includegraphics[width=0.8\linewidth]{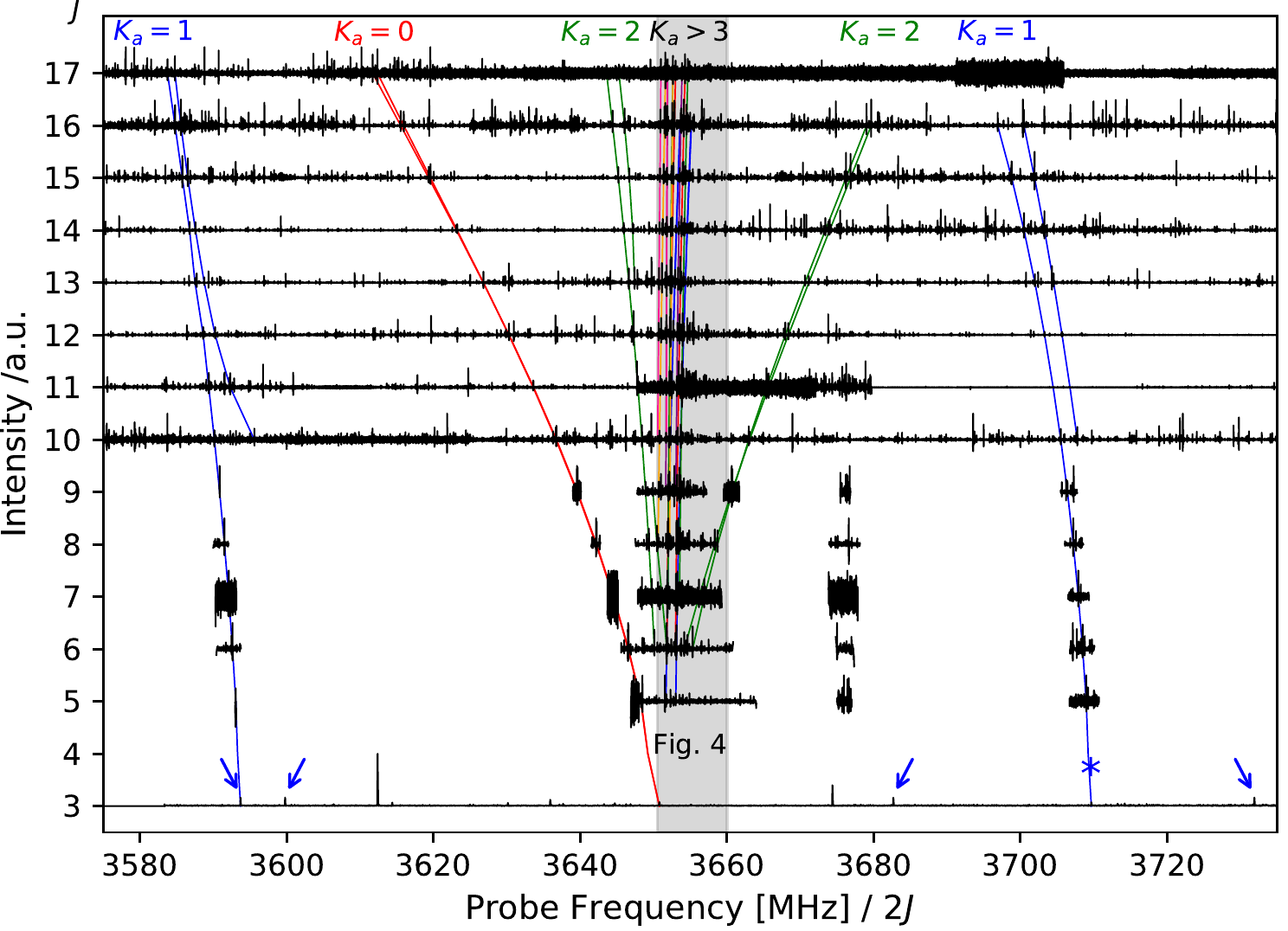} 
\caption{Fortrat diagram of \textit{n}-propanol. The colored lines in the Fortrat diagram mark linkages of $^qR_{K_a}$ series transitions of \textit{Ag n}-propanol, $J_{K_a,K_c}\leftarrow (J-1)_{K_a,K_c-1}$. DM-DR spectroscopy secured assignments in the W-band region (from $J=10$ to $J=16$). Tentative assignments of $K_a=1$ transitions derived from the CP-FTMW measurement (Fig.~\ref{FigA:Propanol_vs_Propanal}) are marked by the blue arrows. Considering found $^qR_{1}$ series, another candidate transition is marked by the blue asterisk. A quantum mechanical model with experimental accuracy can solve the ambiguity of these tentative assignments. The gray area depicts the region of the Fortrat diagram in Fig.~\ref{Fig4:Ag-Fortrat}.  
}
\label{FigA:Fortrat_broad}
\end{figure*}

\begin{figure*}[h]
\centering
\includegraphics[width=0.9\linewidth]{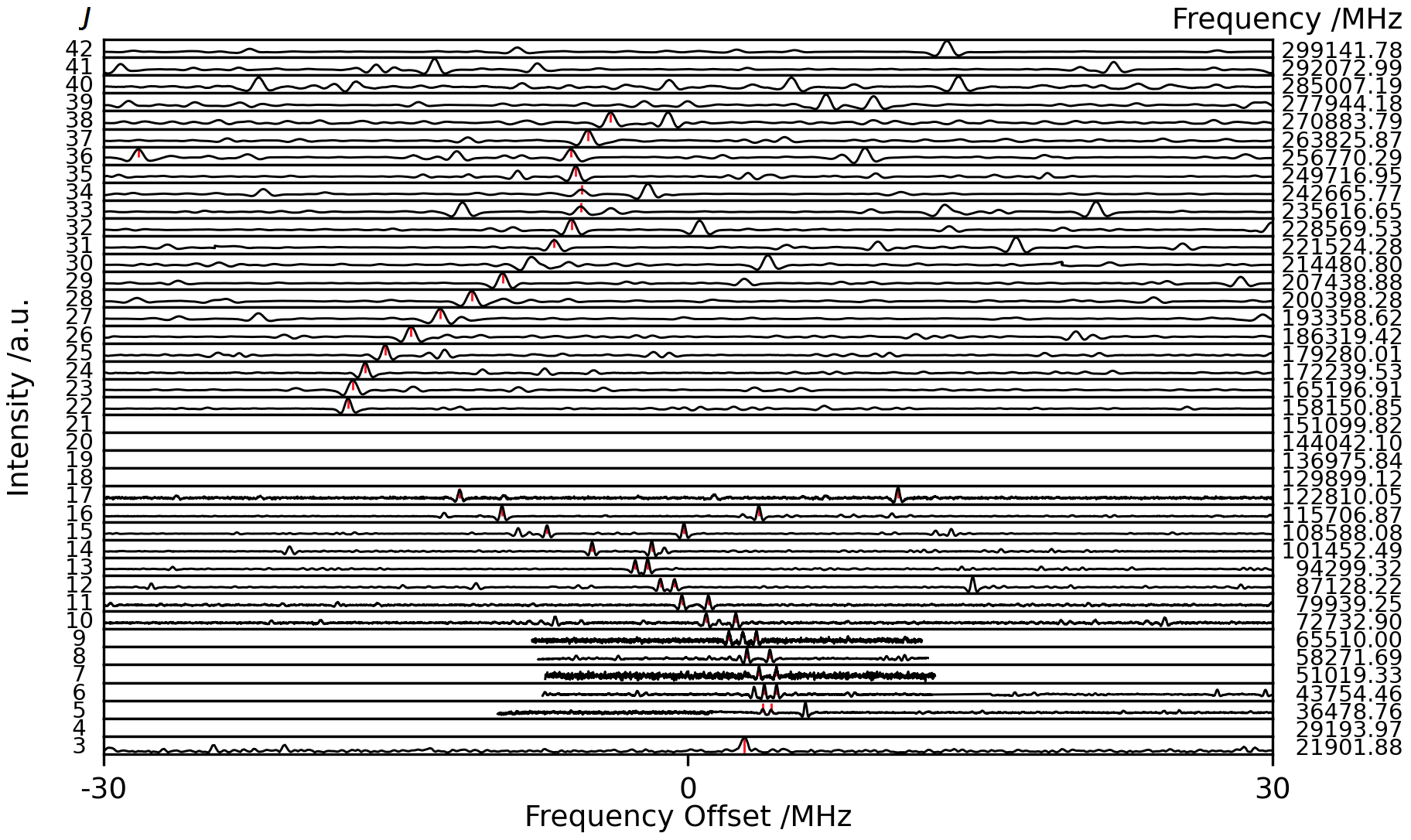} 
\caption{Loomis-Wood plot around the assigned $^qR_0$ series transitions in the W-band region of the \textit{Ag} conformer, which are shown by red sticks.
}
\label{FigA:LW_Ka0}
\end{figure*}

\begin{figure*}[t]
    \begin{subfigure}[b]{\textwidth}
    \includegraphics[width=0.9\linewidth]{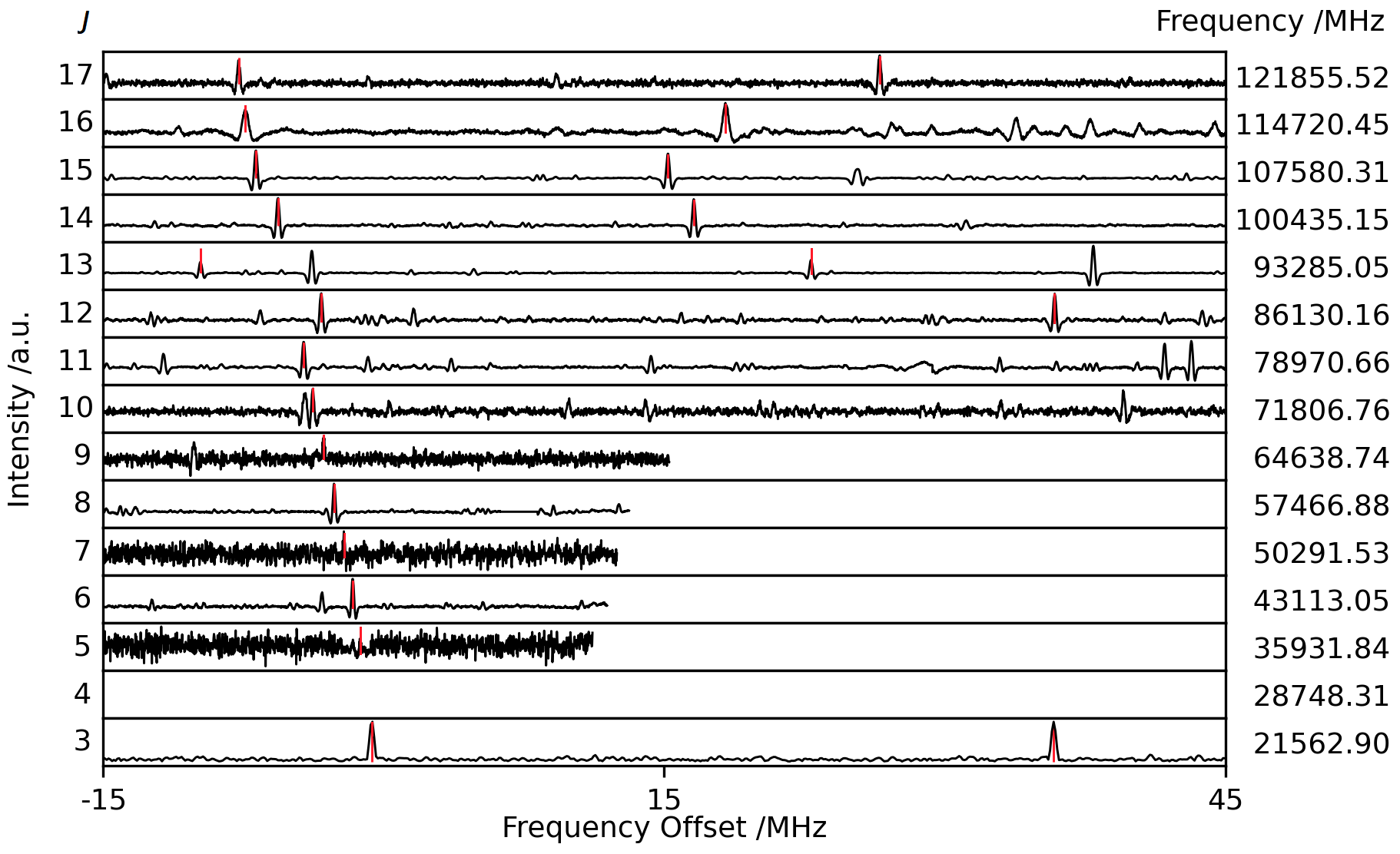}
    \caption{$K_a+K_c=J+1$}
    \end{subfigure}
    \begin{subfigure}[b]{\textwidth}
    \includegraphics[width=0.9\linewidth]{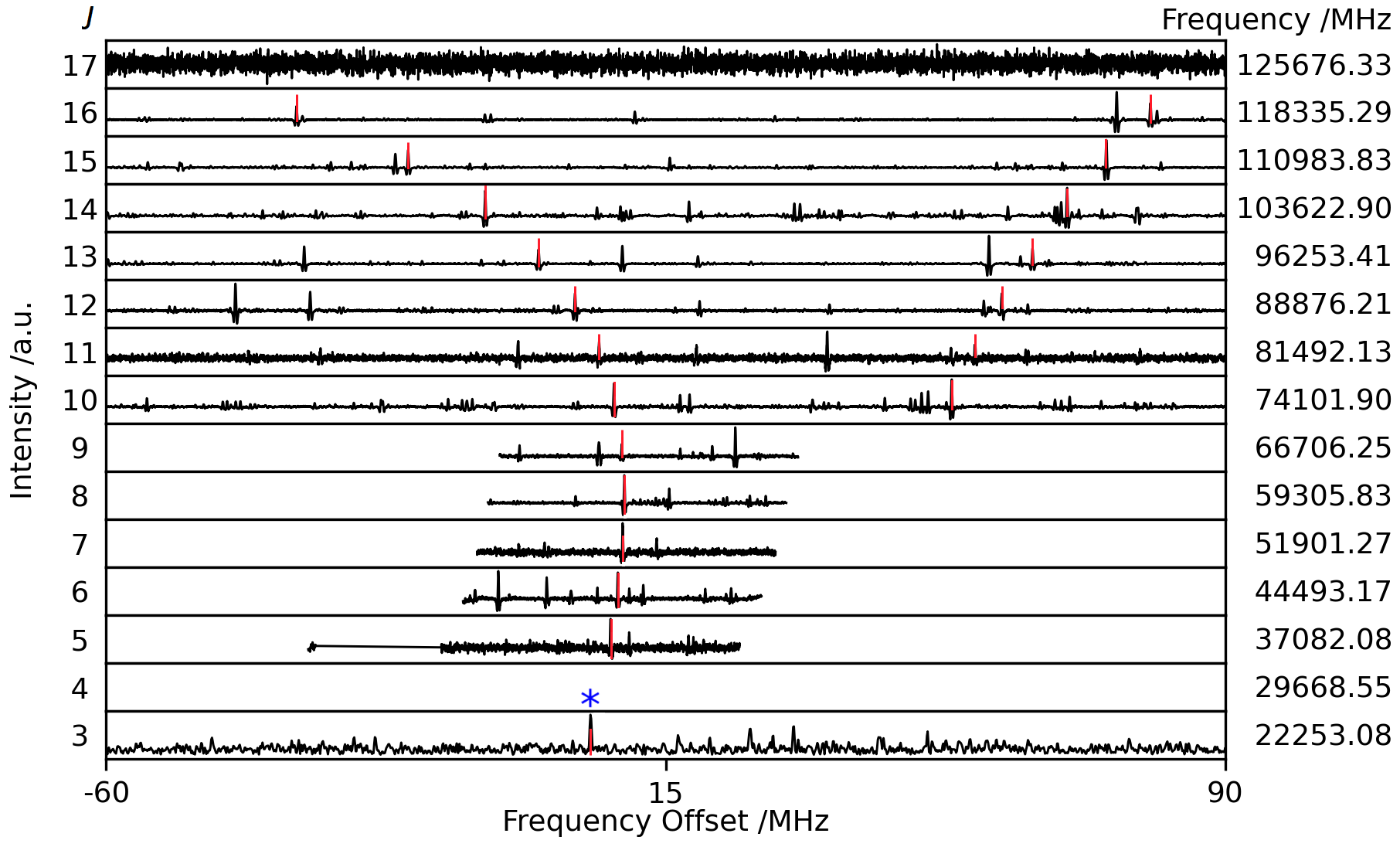} 
    \caption{$K_a+K_c=J$}
    \end{subfigure}
\caption{Loomis-Wood plot around the assigned $^qR_1$ series transitions in the W-band region of the \textit{Ag} conformer, which are shown by red sticks. The blue asterisk (*) marks the same transition as is marked in Fig.~\ref{FigA:Propanol_vs_Propanal}.
}
\label{FigA:LW_Ka1}
\end{figure*}

\begin{figure*}[t]
    \begin{subfigure}[b]{\textwidth}
    \includegraphics[width=0.9\linewidth]{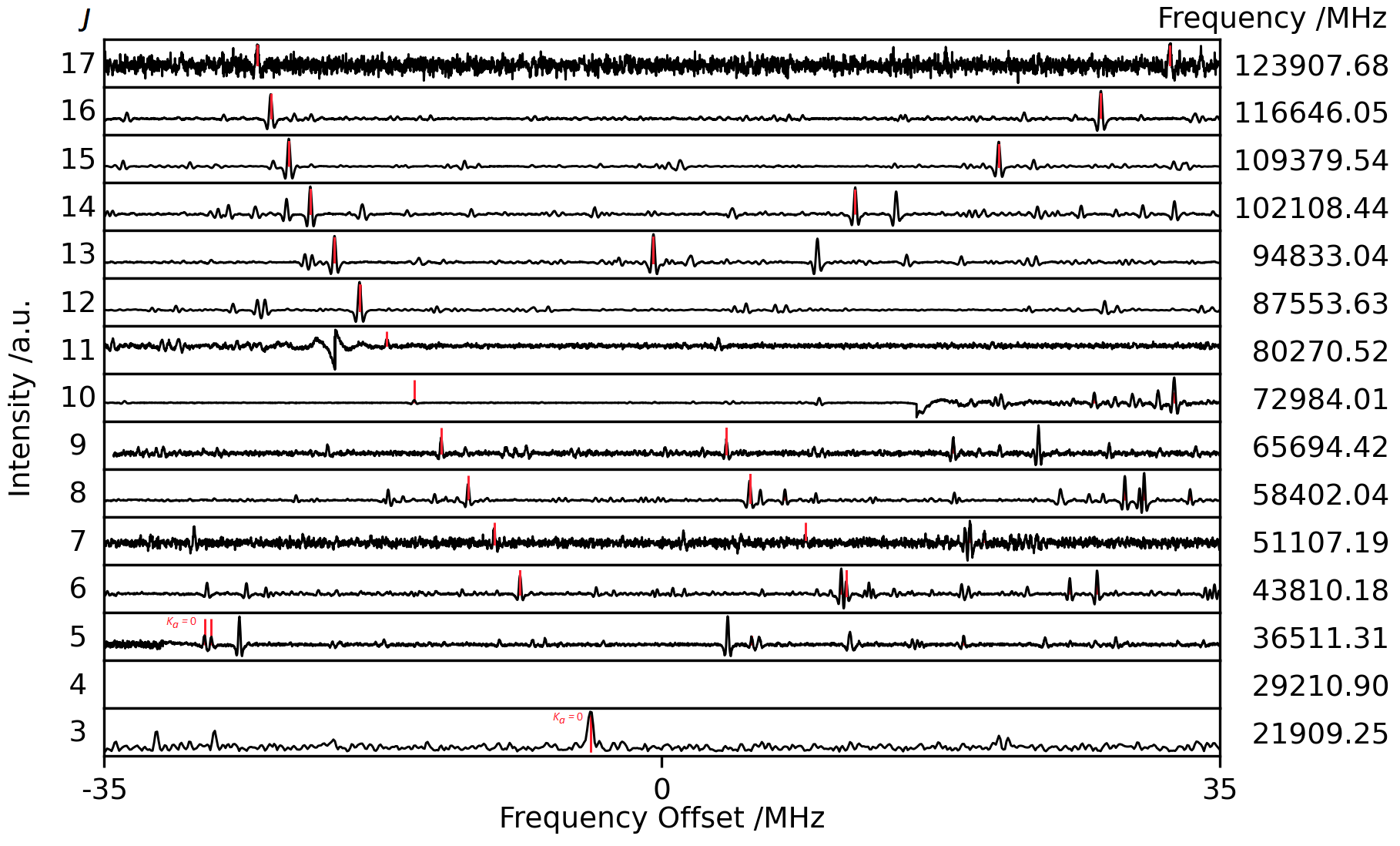}
    \caption{$K_a+K_c=J+1$}
    \end{subfigure}
    \begin{subfigure}[b]{\textwidth}
    \includegraphics[width=0.9\linewidth]{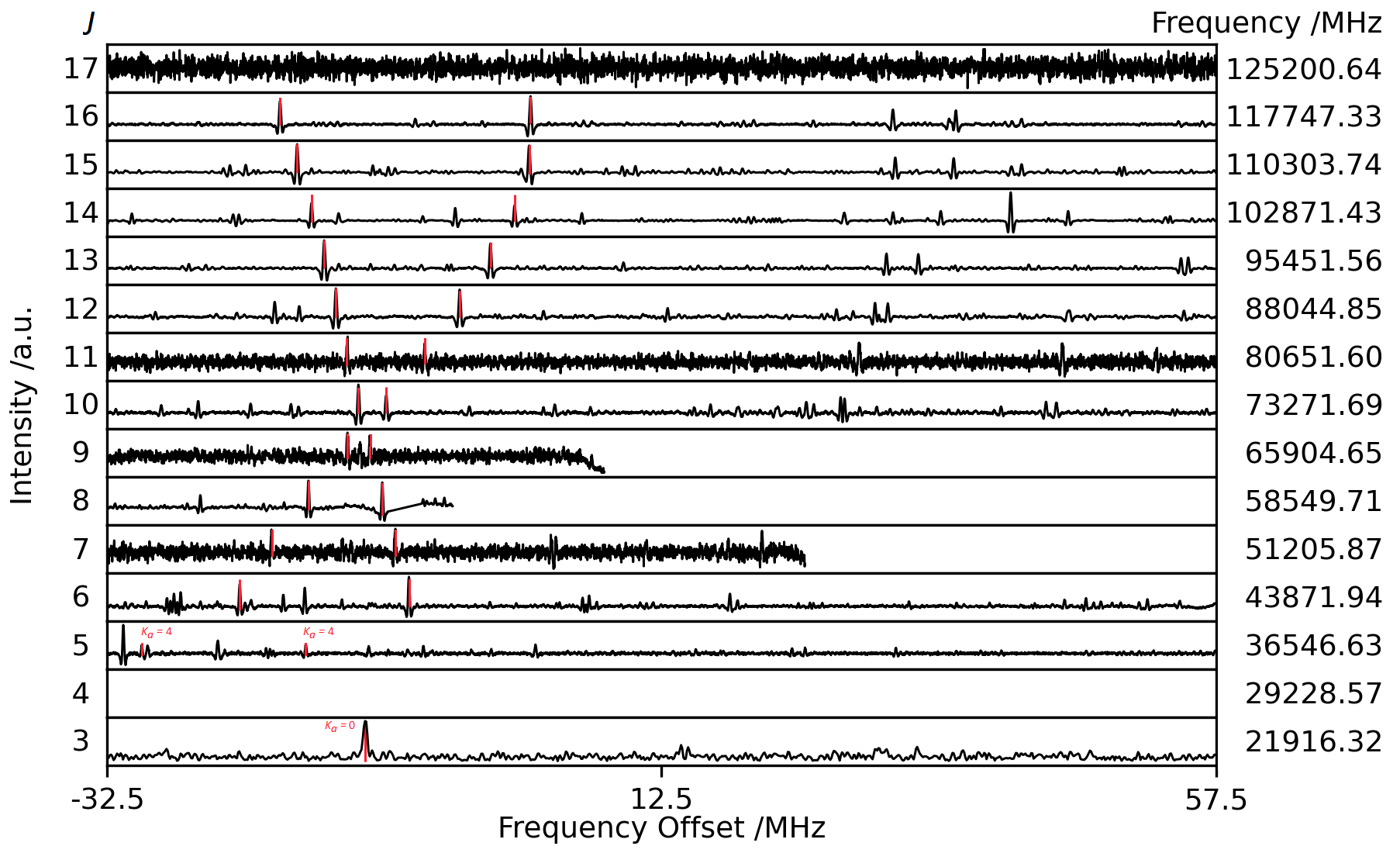} 
    \caption{$K_a+K_c=J$}
    \end{subfigure}
\caption{Loomis-Wood plot around the assigned $^qR_2$ series transitions in the W-band region of the \textit{Ag} conformer, which are shown by red sticks. The series higher in frequency with $K_a+K_c=J+1$ is probably strongly interacting with a so far unknown state around $J=12$. 
}
\label{FigA:LW_Ka2}
\end{figure*}

\begin{figure*}[t]
\includegraphics[width=0.9\linewidth]{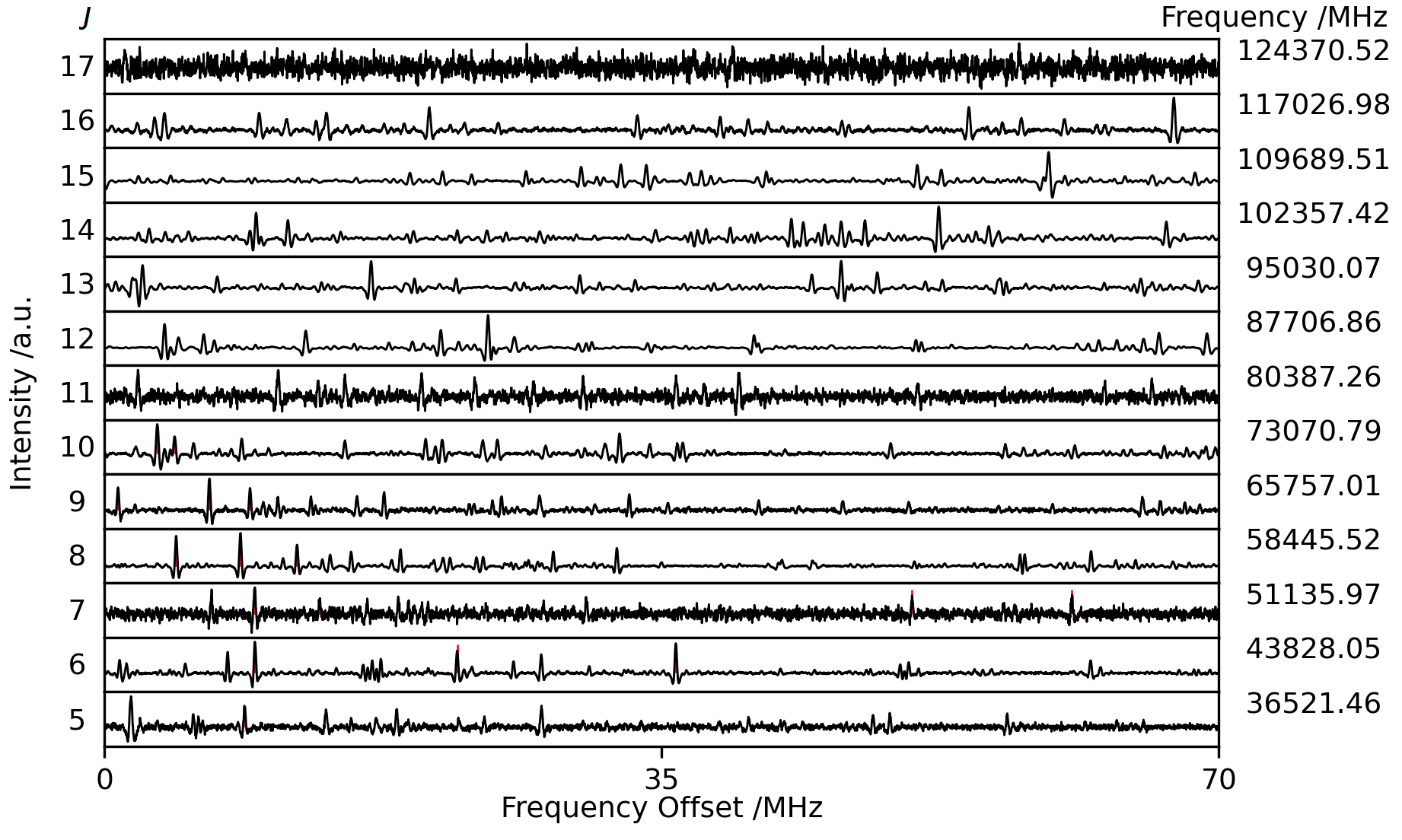}
\caption{Loomis-Wood plot around assumed $^qR_3$ series transitions in the W-band region. No assignments with $K_a=3$ are secured yet, but the existence of such can be seen by the many unassigned transitions in the shown region. Four series close in frequency are expected; Two asymmetry sides for both tunneling states. 
}
\label{FigA:LW_Ka3}
\end{figure*}

First, a part of the MW spectrum of \textit{n}-propanol and propanal, depicting the $3_{K_a,K_c}\leftarrow 2_{K_a,K_c-1}$ $a$-type transitions, recorded with the CP-FTMW spectrometer, are shown in Fig.~\ref{FigA:Propanol_vs_Propanal}.
Rotational signatures are shown in the frequency regions 21.5$-$22.5\,GHz and 25.0$-$26.0\,GHz, respectively. Both measurements were performed under similar conditions.
The rotational spectrum of gauche-propanal is well-understood \citep{ZINGSHEIM2021_Propanal}, in particular doublet structures due to two tunnelings states are observed; Two $K_a=1$ transitions which are close in frequency and have similar intensities.
We note that tunneling-rotation interaction mixes the wave functions of involved rotational energy levels, in this case intensity borrowing may play a role and nominally forbidden transitions can be observed (cf. transitions marked with $\upsilon=0^+\leftrightarrow0^-$). 
The interpretation of the rotational spectrum of \textit{Ag n}-propanol is more complicated. Assigned doublet transitions for \textit{Ag n}-propanol from \citet{Viniti_Ag_1976} are marked in red, but cannot be verified in our CP-FTMW jet experiment. However, two transitions with similar intensities close in frequency are found twice. Comparison of the rotational spectra of both molecules leads to tentative assignments of $K_a=1$ transitions of \textit{Ag n}-propanol. The appearance of two transitions close in frequency were first hints of the existence of two tunneling states ($Ag^+$ and $Ag^-$), similar to what is observed for gauche-propanal. The different tunneling splittings of transitions with $K_a=1$ for \textit{Ag~n}-propanol suggest that involved rotational energy levels are perturbed due to so far untreated interactions. We note that the transition marked with an asterisk (*) is another possible candidate for a transition with $K_a=1$, if $^qR$ series are tracked, cf. Fig.~\ref{FigA:LW_Ka1}. Unambiguously assigning these transitions can be done if a spectroscopic quantum mechanical model with experimental accuracy is derived.\\
Second, typical spectroscopic signatures of $^qR_{K_a}$ series transitions allow for unambiguous assignments of the \textit{Ag} conformer, despite a missing quantum mechanical description.
A broader Fortrat diagram than shown in Fig.~\ref{Fig4:Ag-Fortrat} is presented in Fig.~\ref{FigA:Fortrat_broad}. 
Then, Loomis-Wood plots are shown for series with $0\leq K_a \leq 3$ to further document these assignments (see Figs.~\ref{FigA:LW_Ka0}$-$\ref{FigA:LW_Ka3}). Series with $4\leq K_a \leq 9$, which have been assigned in this work as well, are already well represented by Fig.~\ref{Fig4:Ag-Fortrat} in the main article.
The Loomis-Wood plots are all screenshots of the LLWP software used for assigning transitions within this work \citep{LLWP_Luis}.

\section{Overview of the rotational energy levels of \textit{Aa} and \textit{Ag n}-propanol}

\begin{figure*}[t]
    \begin{subfigure}[h]{\linewidth}
    \centering
    \includegraphics[width=0.80\linewidth]{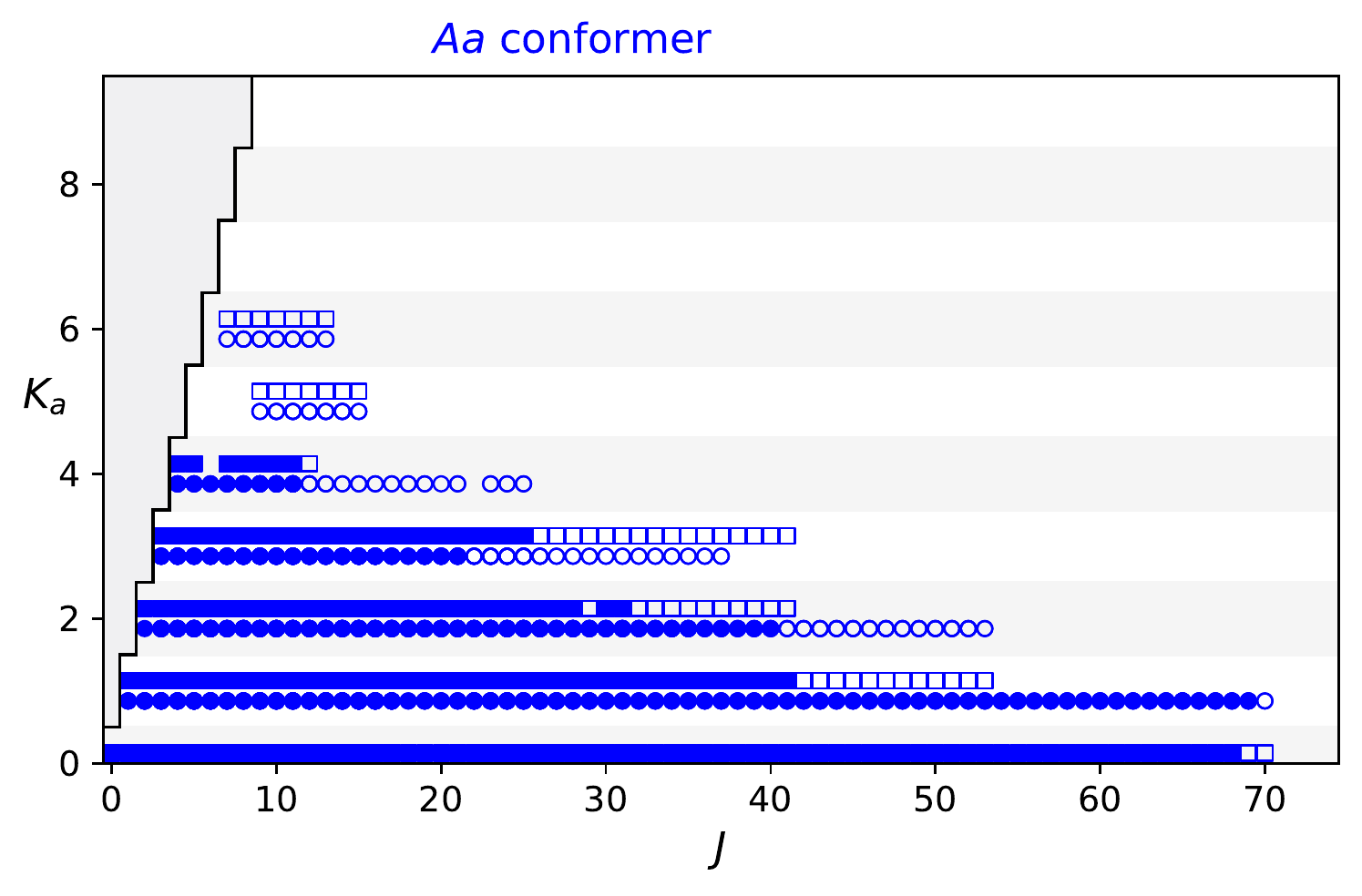} \\
    \caption{\textit{Aa} conformer.}
    \end{subfigure}
    \begin{subfigure}[h]{\linewidth}
    \centering
    \includegraphics[width=0.80\linewidth]{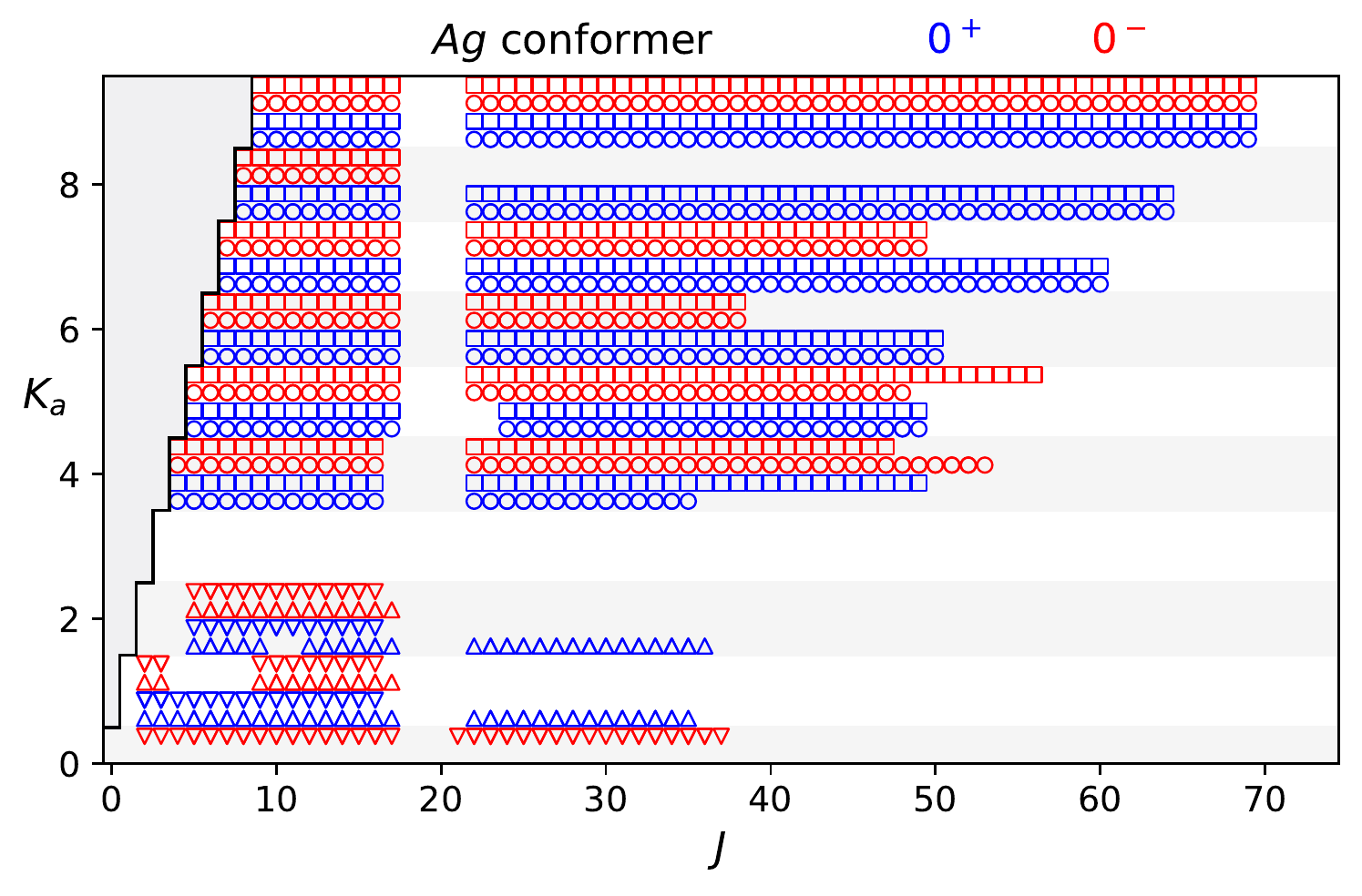}
    \caption{\textit{Ag} conformer.}
    \end{subfigure}
\caption{Quantum number overview of the analysis of (a) \textit{Aa n}-propanol and (b) \textit{Ag n}-propanol.
All rotational energy levels of assigned transitions are visualized with the $J$ quantum numbers on the $x$-axis, $K_a$ quantum numbers on the $y$-axis, and $K_c$ quantum numbers are either depicted by squares or circles for $K_c=J-K_a$ or $K_c=J-K_a+1$, respectively.
Filled markers depict quantum numbers of assigned transitions whose frequencies are fit to experimental accuracy using our final spectroscopic model in Table~\ref{Tab:Spectroscopic_parameters_Aa}. The empty circles mark assigned transitions which are not included in the final analysis, but can be used for astronomical searches.
For \textit{Aa n}-propanol, we note that we do not differentiate between A and E components of each rotational energy level as in overwhelming majority of cases both components are assigned per transition. For \textit{Ag n}-propanol, the rotational energy levels of the $0^+$ and $0^-$ tunneling states are given in blue and red, respectively. Triangles are used for transitions whose tunneling state assignments have only been guessed so far.
}
\label{FigA_QN-overview}
\end{figure*}

\begin{figure*}[t]
    \begin{subfigure}[h]{\linewidth}
    \centering
    \includegraphics[width=0.70\linewidth]{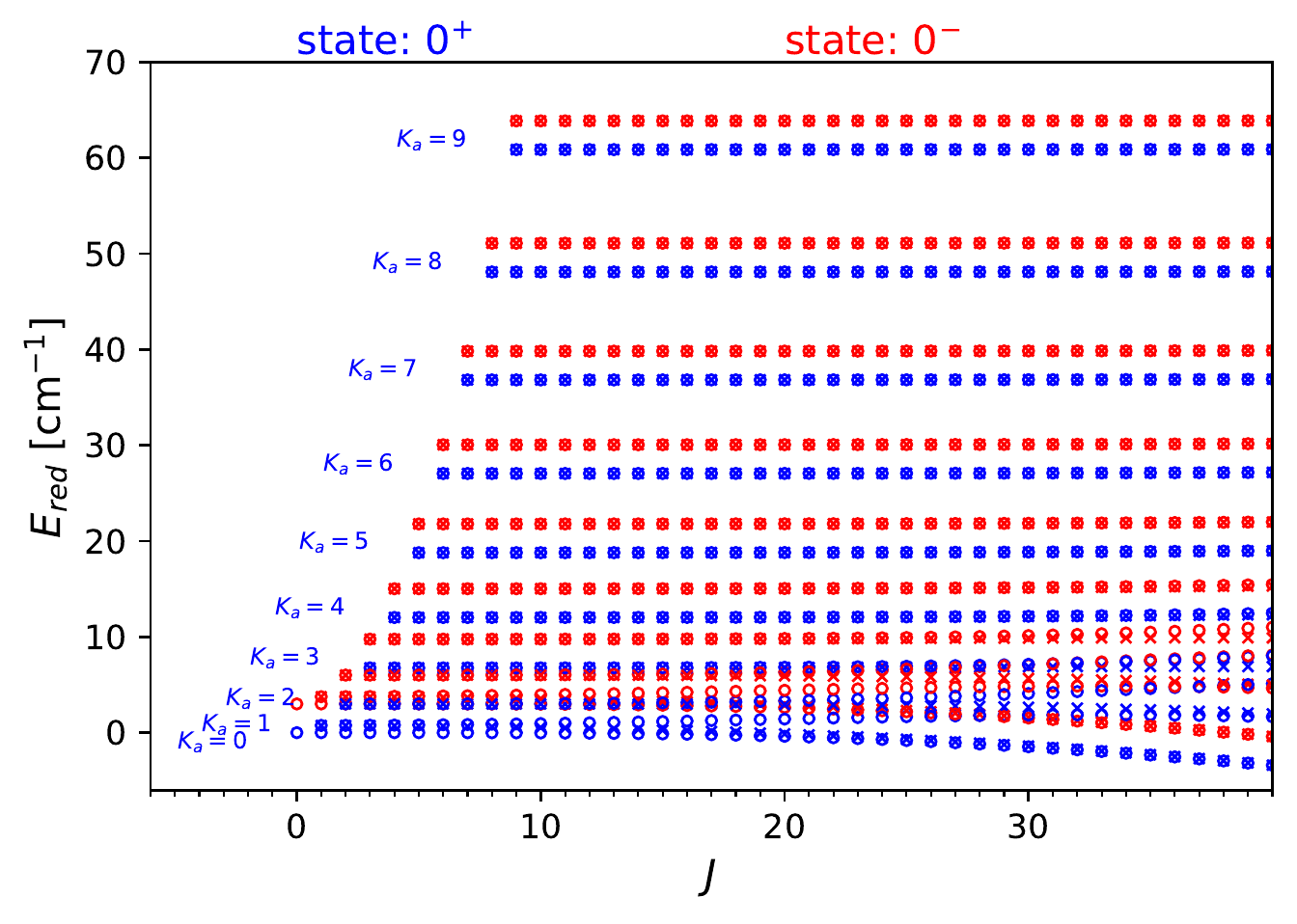} \\
    \caption{Reduced energy diagram for rotational energy levels with $K_a\leq9$.}
    \end{subfigure}
    \begin{subfigure}[h]{\linewidth}
    \centering
    \includegraphics[width=0.70\linewidth]{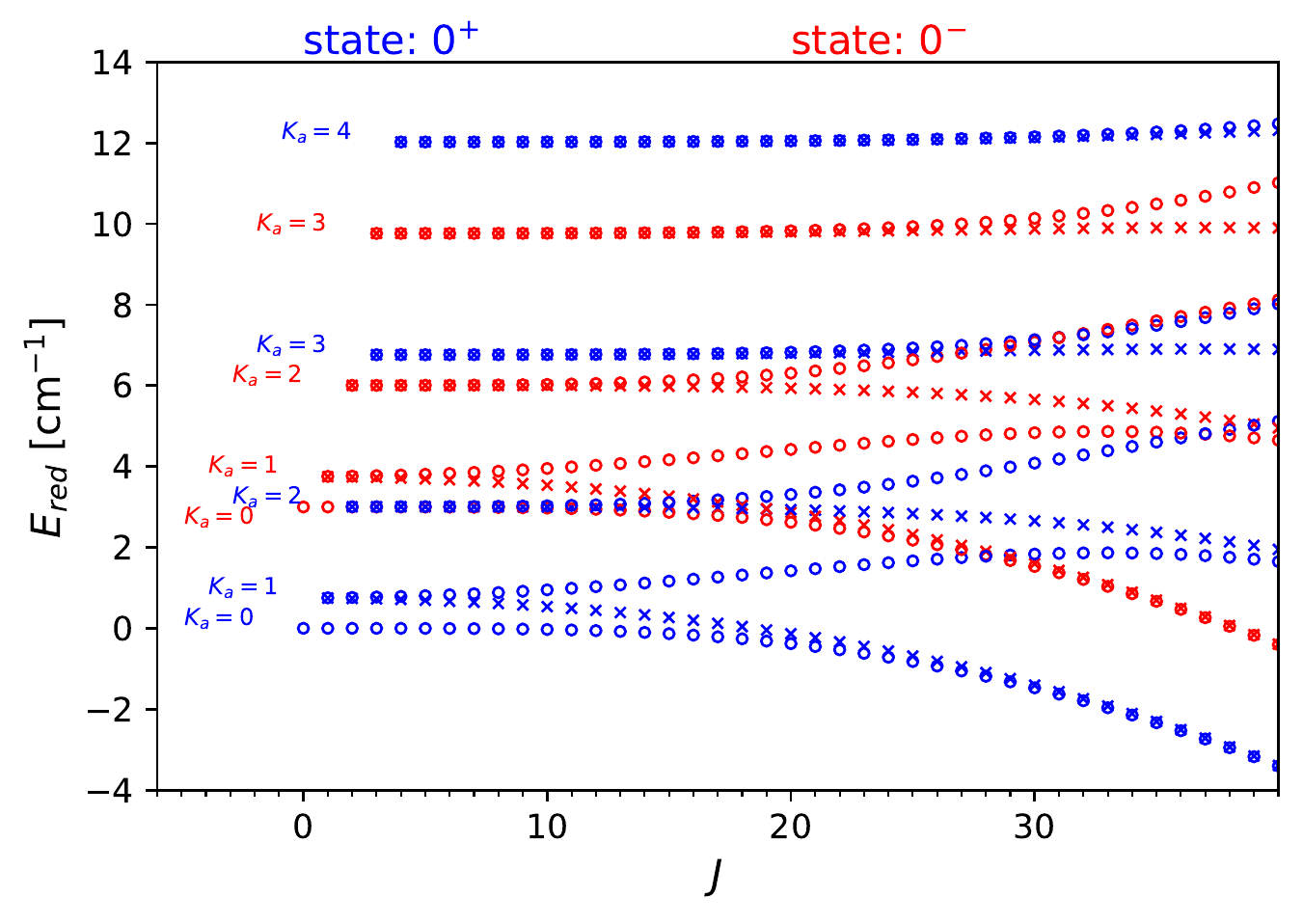}
    \caption{Reduced energy diagram for rotational energy levels with $K_a\leq3$.}
    \end{subfigure}
\caption{Reduced energy diagram of \textit{Ag n}-propanol.
The reduced energy is $E_{\textrm{red}}= E_{\textrm{T}}+E_{\textrm{rot}} - J \cdot (J+1) \cdot (B+C)/2$.
Thereby, the energies of the rotational levels $E_{\textrm{rot}}$ are calculated based on ab initio calculations of \citet{n-propanol_Kisiel_2010}. We used the same parameters for both tunneling states as  ab initio calculations are presented for a single state only.
We note that actual parameters of the two tunneling states will differ.
We set the energy difference of the two tunneling states to 3\,cm$^{-1}$, i.e. $E_{\textrm{T}}^{0^+}$=0\,cm$^{-1}$ and $E_{\textrm{T}}^{0^-}$=3\,cm$^{-1}$, on the assumption that the splitting is similar to what has been observed for \textit{gauche}-ethanol \citep{ethanol_gauche_1997}. 
\textbf{(a)} The reduced energy diagram for rotational energy levels with $K_a<9$ shows that series with $K_a>5$ may not cross each other. We note that it may happen if different rotational parameters for the two tunneling states are taken into account.
\textbf{(b)} The reduced energy diagram for rotational energy levels with $K_a\leq4$ is a zoom of the $y$-axis of (a).
We note that many series cross each other and even though the exact energy difference between the two tunneling states $Ag^+$ and $Ag^-$ is unknown so far, many tunneling-rotation interactions occur in this region and determining an accurate energy difference is essential to properly take them into account.
}
\label{FigA_red-egy}
\end{figure*}

We present a quantum number overview of our analysis for \textit{Aa} and \textit{Ag} conformers of \textit{n}-propanol here as we could not reproduce all assigned transitions to experimental uncertainty. However, all assigned transitions can be used for astronomical searches of the rotational spectra of both conformers, therefore, quantum number overviews are given for clarification (see Fig.~\ref{FigA_QN-overview}).
Furthermore, an estimated reduced energy diagram of \textit{Ag n}-propanol is shown, which illustrates the location of tunneling-rotation interactions (see Fig.~\ref{FigA_red-egy}). Of course, interactions between rotational energy levels of \textit{Ag n}-propanol and those of \textit{Aa n}-propanol should also be considered if a global analysis is performed, but these interactions are not illustrated here.

\end{appendix}
\end{document}